\documentclass[showpacs,preprintnumbers,nofootinbib,amsmath,amssymb]{revtex4-2}

\usepackage{color,graphics,epsfig,graphpap}

\usepackage{graphicx}
\usepackage{dcolumn}
\usepackage{bm}

\newcommand{\be}{\begin{eqnarray}}
 \newcommand{\ee}{\end{eqnarray}}
 \newcommand{\nee}{\nonumber\end{eqnarray}}
 \newcommand{\nn}{\nonumber\\}
  \newcommand{\wt}{\widetilde}

\def\a{\alpha}
\def\s{\sigma}

\begin{document}

\author{D. Kotlorz\footnote{On leave of absence from Department of Physics, Opole University
of Technology, 45-758 Opole, Proszkowska 76, Poland}}
\email {dorota@theor.jinr.ru} \affiliation{Bogoliubov Laboratory of Theoretical Physics,
JINR, 141980 Dubna, Russia}
\author{E. Christova}
\email {echristo@inrne.bas.bg} \affiliation{Institute for Nuclear Research and Nuclear Energy, Bulgarian Academy of Sciences,
Tzarigradsko chauss\'{e}e 72, 1784 Sofia, Bulgaria}
\author{E. Leader}
\email {e.leader@imperial.ac.uk} \affiliation{ Imperial College London, London SW7 2AZ, United Kingdom}

\title{Tests of the parametrizations of Fragmentation Functions using data on
inclusive pion and kaon production in unpolarized $pp$ collisions
from the STAR collaboration and at the NICA project.
}

\begin{abstract}
The goal of this study is to check which, if any, of the published versions
of the pion and kaon fragmentation functions is compatible with the STAR data on
semi-inclusive pion and kaon production in proton-proton collisions, and on the basis
of this analysis to make reliable predictions for the $p_T$ spectra of the pions
and kaons in inclusive pion and kaon production at the future NICA proton-proton collider.

The calculations are carried out in next-to-leading order (NLO) of perturbative
quantum chromodynamics (pQCD), using the well tested CTEQ6 parton distributions.
We consider the following pion and kaon fragmentation functions (FFs) -- 
DSEHS-14~\cite{deFlorian:2014xna}, DSEHS-17~\cite{deFlorian:2017lwf},
LSS-15~\cite{Leader:2015hna}, HKNS-07~\cite{Hirai:2007cx}
and AKK-08 \cite{Albino:2008fy}. Our analysis shows that within the experimental
errors all tested sets of fragmentation functions provide a good fit to STAR
data at the c.m. energy $\sqrt{S} = 200\,{\rm GeV}$, and the best ones are
both LSS-15 and DSEHS-14 for pions and DSEHS-17 for kaons.
From comparison of the LO and NLO results it is clear that the latter fit
data much better, specially in the region of small $p_T$. The NLO cross sections are also less
scale-$Q^2$ dependent, where $p_T/2\leqslant Q \leqslant 2p_T$, than the LO ones.

In order to make predictions for NICA energies, we compare the NLO pQCD results with the
existing experimental BES STAR data on semi-inclusive hadron production in
the most peripheral Au+Au collisions where the nuclear effects can be neglected.
The comparison for lower energy scales, like at NICA, shows
that a purely pQCD approach is inadequate and suggests the necessity to take into account
also higher-order effects of initial-state soft-gluon radiation.
Nevertheless, these data on the $p_T$ spectra of $\pi^+$, $K^+$ and also the ratios
$\pi^-/\pi^+$ and $K^-/K^+$ seem favour LSS-15 and DSEHS-14 FFs for pions
and DSEHS-17 for kaons, similarly as at the energy scale $\sqrt{S} = 200\,{\rm GeV}$.
\end{abstract}

\date{\today}

\maketitle

\section{Introduction}

Study of the spectra of identified hadrons at high $p_T$ in $pp$
collisions is based on model calculations in perturbative quantum chromodynamics (pQCD)
\cite{Jager:2002xm}. Both in leading order (LO) \cite{Combridge:1977dm, Cutler:1977qm,
Owens:1977sj, Feynman:1978dt, Baier:1979tp} and in next-to-leading order (NLO)
\cite{Ellis:1979sj, Ellis:1985er, Aversa:1988vb} of pQCD calculations, inclusive
production of single hadrons is described in terms of parton distribution
functions (PDFs), parton-parton interaction cross sections calculated in the Standard Model (SM),
and fragmentation functions. The PDFs are fairly well known, while on the
contrary, the flavor-separated quark and gluon FFs, being relatively new
objects, required for a quantitative description of hard scattering processes
involving identified light hadrons in the final-state,
are not so well constrained. At present  there are several sets of FFs that fairly well describe data, but nevertheless differ quite a lot in different kinematic regions. 
Most directly the FFs have been extracted from one-hadron production in electron-positron collisions.
However, this process in principle, cannot distinguish the quark and anti-quark FFs and  information only
about $D^h_{q + \bar q}$ is obtained. In order to obtain separate quark and anti-quark FFs, one-hadron
semi-inclusive $lN$ and $pp$ processes play an essential role.
However, as the hadron structure enters in comparing theoretical calculations with experimental data, different model assumptions have to be made.
One of the goals in the experimental measurements by the STAR collaboration of
pions and kaons at high $p_T$ in $p p$ collisions is  to provide more information on the FFs.

In this paper, based fully on pQCD and taking into account all partonic cross sections in NLO, we calculate
the hadronic $p_T$-spectra of the final pions and kaons in single hadron production in $pp$ collisions
in the high $p_T$ region, and compare it to the STAR data from 2006 \cite{STAR:2006xud},
2007 \cite{STAR:2006nmo} and 2012 \cite{Agakishiev:2011dc}.
We use various published parametrizations for the FFs: HKNS-07 \cite{Hirai:2007cx},
AKK-08 \cite{Albino:2008fy}, DSEHS-14~\cite{deFlorian:2014xna}, DSEHS-17~\cite{deFlorian:2017lwf}
and LSS-15 \cite{Leader:2015hna}, aiming to obtain constraints on the available sets of FFs from the
sensitivity of STAR data to the used FFs.
We use CTEQ6 parametrization for PDFs \cite{Pumplin:2002vw}.
Using these data to constrain the FFs, should lead to more reliable predictions for future
measurements. In particular we consider the effect on data expected from the future NICA accelerator.

Our analysis is partially based on the model independent approach of the difference cross sections of $h^+$ and $h^-$-production, developed in
\cite{Christova:2008te}.
It suggests that, both in $e^+e^-$, SIDIS and proton-proton collisions, if instead of the X-sections $d\s_{e^+e^-}^h$, $d\sigma_N^h$ and $d\sigma_{pp}^h$  for inclusive
production of hadrons and their antiparticles, one deals with
their differences, i.e. with $d\s_{e^+e^-}^{h-\bar h}$, $d\s_N^{h-\bar h}$ and
$d\sigma_{pp}^{h-\bar h}\equiv d\sigma_{pp}^h-d\sigma_{pp}^{\bar h}$,  one determines directly
the non-singlet (NS) combinations of FFs $D_q^{h-\bar h}=D_{q_V}^h$.
This result follows solely from charge-conjugation invariance of the strong interactions without any
model assumptions about the sea-quarks or favoured and unfavoured fragmentation functions.

The STAR Collaboration has presented data for pions and kaons. In comparing with the data we make
two simplifications in our formula for the spectra of the pions and the kaons.
1) We use $s=\bar s$ for the strange quark PDFs. Note that this is not an assumption but follows from
the strong limit $\vert s-\bar s\vert <0.025$ obtained in neutrino experiments \cite{Bourrely:2007if}
and implies that the contribution from terms proportional to $s-\bar s$ will be within the experimental
errors and thus negligible.
2) We assume SU(2) isospin symmetry of the strong interactions.
The theoretical framework for inclusive production of single hadrons in $pp$
collisions is presented in Section II, and the expressions for the difference cross sections and
the conditions under which they have been derived are given in Section III.
The comparison of our results to the STAR data is presented in Section IV.
Here, we also compare the results obtained within NLO and LO approaches and
study the scale dependence of the cross sections for the charged hadrons at LO and NLO where
the scale $Q$ may vary from $p_T/2$ to $2p_T$.

Unfortunately we cannot take advantage of the very simple expressions derived in Section III
for the cross-section differences, because the experimental errors on these are too large.

Based on the obtained results for the FFs, we calculate the expected $p_T$-spectra of the
pions and kaons produced in $pp$ collisions at the kinematics of the planned NICA accelerator.
We compare the theoretical NLO pQCD predictions with the experimental BES STAR
data on semi-inclusive pion and kaon production in the most peripheral Au+Au
collisions, where nuclear effects can be neglected
\cite{PHENIX:2001hpc,STAR:2003jwm}, at energies $\sqrt{S} = 11.5$ and 27~GeV,
specific to NICA \cite{Kekelidze:2020hdy,Taranenko:2020vqn}, and show that
at such low energy scales and the correspondingly low values of $p_T$,
the contributions to the inclusive cross section for hadron production
coming from various soft processes become essential. Therefore,
the perturbative QCD approach should be here treated with caution.
This is discussed in Section V.
We summarize the obtained results in a Conclusion.

\section{The theoretical framework for $pp\to h+X$}

We consider the process:
\be
p(P_A)+p(P_B)\to h(P^h)+X\label{1}
\ee

The basic concept \cite{Jager:2002xm} underlying the theoretical analysis of
most high energy interactions is the factorization theorem, which states that
the cross-section for  large momentum-transfer reactions may be factorized into
long-distance pieces that contain the desired information on the structure of
the nucleon in terms of its parton densities such as $q(x )$ and fragmentation
functions $D^h_q(z)$, and short-distance parts which describe
the hard interactions of the partons.
The strength of this interaction is controlled by the running strong-interaction
coupling constant $\alpha_s$ evaluated at a large scale associated with the
hard interaction which we denote by $Q^2$. The effective coupling constant
falls logarithmically with increasing $Q^2$,
$\alpha_s(Q^2)\sim 1/\ln(Q^2/\Lambda^2_{\rm QCD})$, enabling the perturbative QCD
analysis for the hard-scale $Q\gg \Lambda_{\rm QCD}$, where
$\Lambda_{\rm QCD}\sim 0.3\, {\rm GeV}$.
The two crucial points here are, on one hand
the long-distance contributions are universal, i.e., they are the same in any inelastic reaction
under consideration, and  on the other hand the short-distance pieces depend only on the
large scales related to the large momentum transfer in the overall reaction and, therefore, can
be evaluated using QCD perturbation theory.

Calculations in  LO, Eq.~(\ref{1}), have been considered and compared to data in several papers in
\cite{Combridge:1977dm, Cutler:1977qm, Owens:1977sj, Feynman:1978dt, Baier:1979tp}.
The first calculations beyond LO, at $(\alpha_s^3)$ order in perturbative QCD, that include the virtual corrections in the $2\to 2$ and the $2\to 3$ partonic subprocesses,
have been calculated in \cite{Ellis:1979sj, Ellis:1985er, Aversa:1988vb}, and the full
$\mathcal{O}(\alpha_s^3)$ radiative NLO corrections are presented in \cite{Aversa:1988vb}.

The results presented for the unpolarized cross section for $ pp \to \pi +X$
at $\sqrt{S} = 200$ GeV, that are well described by
the NLO QCD calculation \cite{Torii:2002tq}, provide confidence that the theoretical framework based on
perturbative-QCD hard scattering  is adequate.\\

Here we consider cross sections $ pp \to h^{\pm} +X$ and also
difference cross sections $\sigma(pp\to h^+ +X) - \sigma(pp\to h^- +X)$.
The calculations are done in NLO in the QCD improved parton model
following the approach of \cite{Aversa:1988vb} which include all parton subprocesses involving quarks and gluons.
In our numerical analysis, we have adapted the INCNLO inclusive hadron production code
\cite{INCNLO} and also a code for calculating the fragmentation functions \cite{Hirai:2007cx}.

In the simple parton model process (\ref{1}) proceeds via the $2\to 2$ partonic subprocesses:
\be
q_a(p_a)+q_b(p_b)\to c(p_c)+d(p_d)\label{2}
\ee
where  $q_a, q_b$ and $c,d$ can be either quarks or gluons.
In what follows all kinematic variables refer to the CM of the $pp$ collision.
The expression for the cross section for $pp\to hX$ in the c.m.s. of $pp$ has the factorized form:
\be
E^h\frac{d\sigma_{pp}^{h}}{d^3P^h} &=&\frac{1}{\pi}\,\sum_{ab\to cd}\int^1_{x_{a,min}}dx_a
\int^1_{x_{b,min}}dx_b\, \frac{1}{z }\,\times \nn
&&\times \left\{
q_a(x_a)q_b(x_b)\,\left[\frac{d\hat\sigma_{ab}^{cd}}{dt}\,D_c^h(z )+\frac{d\hat\sigma_{ab}^{cd}}{du}\,D_d^h(z )\right]
+q_a(x_b)q_b(x_a)\,\left[\frac{d\hat\sigma_{ab}^{cd}}{du}\,D_c^h(z )+\frac{d\hat\sigma_{ab}^{cd}}{dt}\,D_d^h(z )\right]\right\}.
\label{Ks}
\ee
where $d\hat\sigma_{ab}^{cd}$ are the Born cross sections to order $\a^2_s$ and  the parton densities and
FFs are scale independent,  $z =E^h/E_c$, and where
\be
\frac{d\hat\sigma_{ab}^{cd}}{dt}\equiv \frac{d\hat\sigma_{ab}^{cd}}{dt}\,(s,t,u),\qquad \frac{d\hat\sigma_{ab}^{cd}}{du}\equiv \frac{d\hat\sigma_{ab}^{cd}}{dt}\,(s,u,t).
\ee
As usual $x_a$ ($x_b$) is the fraction of the proton momentum $P_A$ ($P_B$) carried by the parton $q_a$ ($q_b$), collinear to the momentum of the initial hadron $A$ ($B$).

In the QCD improved parton model, when QCD corrections are included, the factorized
form of (\ref{Ks}) is preserved and the PDFs and FFs are replaced by $Q^2$-dependent distribution functions:
\be
q(x) \to q(x, Q^2),\qquad D_q^h(z) \to D_q^h(z, Q^2),
\ee
where $Q^2$ is some relevant large momentum scale.
Furthermore, the effective quark-gluon-quark coupling constant $\a_s$ becomes
the running coupling given at NLO by
\be
\a_s(Q^2)=\frac{4\pi}{\beta_0 \ln (Q^2/\Lambda^2_{\rm QCD} )}
\left (1-\frac{\beta_1\ln \ln (Q^2/\Lambda^2_{\rm QCD})}
{\beta_0^2 \ln (Q^2/\Lambda^2_{\rm QCD} )}\right ),
\ee
where $\beta_0=11-2/3\,N_f$ for $N_f$ flavours, $\beta_1=102-38/3\, N_f$ and
NLO $\Lambda_{\rm QCD}=0.32\, {\rm GeV}$.
The momentum scale $Q^2$ is set to the ``natural'' scale $Q^2=p_T^2$
\cite{Ellis:1985er}.
In order to use the PDFs and FFs, known from some other process at some other
``input'' scale $Q_0^2=1\, {\rm GeV^2}$,
we evolve the PDFs and FFs from $Q_0^2$ to $Q^2>Q_0^2$ via the NLO
DGLAP-equations. This procedure works well for hard-scale
$Q\gg \Lambda_{\rm QCD}$, typically for $Q^2$ above $\sim 1\,{\rm GeV^2}$,
where the short-distance part of the strong interactions is expected to
dominate and hence the perturbative QCD methods can be applied.

In our approach when using the difference cross sections, only NS
combinations of  FFs enter -- see for example Eqs. (\ref{sh}) and (\ref{Kp}).
In this case the DGLAP evolution equations for a given FF is especially simple
with no mixing with other FFs.

The  Mandelstam variables of the partonic process (\ref{2}) are:
\be
\label{partXsections}
  s&=&(p_a+p_b)^2= (x_aP_A + x_bP_B)^2=x_a\,x_b\,S\nn
t&=&(p_a-p_c)^2= (x_aP_A -\frac{P^h}{z })^2=\frac{x_a}{z }\,T,\nn
u&=&(p_b-p_c)^2= (x_bP_B - \frac{P^h}{z })^2=\frac{x_b}{z }\,U
\ee
Here  $c$ or $d$ are  the fragmenting partons
that are assumed collinear to the final hadron $h$, and
everywhere masses are neglected, so that $s+t+u=0$ holds,  which determines the value of $z $:
  \be
z =-\,\frac{x_a\,T+x_b\,U}{x_a\,x_b\,S}
  \ee
    The letters  $S,\,T$ and $U$ stand for the Mandelstam variables of the inclusive  hadronic process (\ref{1}):
\be
S&=& (P_A+P_B)^2=2\,(P_A\cdot P_B)=4E^2\\
T&=& (P_A-P^h)^2=-\,2\,(P_A\cdot P^h)=-2EE^h(1-\cos\theta )\\
U&=& (P_B-P^h)^2=-\,2\,(P_B\cdot P^h)=-2EE^h(1+\cos\theta )\\
p_T^2&=&\frac{UT}{S}=(E^h)^2\sin^2\theta\, ,
\ee
where, as mentioned, the kinematic variables refer to the CM of the $pp$
collision; $E$ is the energy of the colliding proton beams.
The maximal value of the  transverse momentum of a hadron $h$  produced with energy $E^h$ in $pp $ collisions is
\be
p_T^{max}&=&E^h, \,\,\textrm{at} \,\, \theta =\pi/2\, .
\ee
Often the $p_T$-spectra are presented in terms of  its relative  value $x_T=2p_T/\sqrt S$, where $\sqrt S/2=E$.

The lower limits of integration are determined by the conditions $z <1$ and $s+t+u=0$
\cite{Owens:1977sj}:
\be
x_{a,min}&=&\frac{-U}{T+S} \\
x_{b,min}&=&\frac{-x_aT}{x_aS+U}
\ee

There are 8 different $2\to 2$ partonic processes that contribute to $d\s^h_{pp}$:
\be
\hat \sigma_1:\,&&q_iq_j\to q_iq_j,\quad \bar q_i\bar q_j\to \bar q_i \bar q_j,\quad q_i\bar q_j\to q_i\bar q_j, i\neq j\nn
\hat\sigma_2:\,&&q_iq_i\to q_iq_i,\quad \bar q_i\bar q_i\to \bar q_i \bar q_i,\nn
\hat\sigma_3:\,&&q_i\bar q_i\to q_j\bar q_j,\quad i\neq j\nn
\hat\sigma_4:\, &&q_i\bar q_i\to q_i\bar q_i\nn
\hat\sigma_5:\,&& q_i\bar q_i\to gg\nn
\hat\sigma_6:\,&& gg\to q_i\bar q_i\nn
\hat\sigma_7:\, &&q_ig\to q_i g\nn
\hat\sigma_8:\,&& gg\to gg
\ee
expressions for which can be found in many places.
In NLO $\mathcal{O}(\alpha_s^3)$, additional $2\to 3$ scattering processes
also contribute \cite{Ellis:1979sj, Ellis:1985er, Aversa:1988vb}.

The contributions of the partonic processes which occur in Eq.~(\ref{1}) are given in Appendix A.

\section{The difference cross sections : $h^+-h^-$}

Here we would like to introduce the idea of the difference cross sections
and discuss its possible advantages.
We define the difference cross section:
 \be
\sigma_{pp}^{h^+-h^-} \equiv
\sigma_{pp}^{h^+}-\sigma_{pp}^{h^-}.
\ee
C-invariance of strong interactions implies:
 \be
 &&D_g^{h^+-h^-}=0,\label{Dg}\\
&&D_q^{h^+-h^-}=-D_{\bar q}^{h^+-h^-}=D_{q_V}^{h^+}\label{Dq}.
\ee
The cross section $\sigma_{pp}^{h^+-h^-}$ has especially simply form in LO,
where, without any assumptions about the sea quark PDFs or the favoured and
unfavoured FFs, we obtain:
\be
&&E^h\frac{d\,\sigma_{pp}^{h^+-h^-}}{d^3P^h} =
\frac{1}{\pi}\int dx_a\,dx_b\,\frac{1}{z }\sum_{q=u,d,s}D_{q_V}^{h^+}(z )
\big( q_V(x_a)\,L_q(x_b,t,u)+q_V(x_b)L_q(x_a,u,t)\big)\label{sh},
\ee
where
\be
L_u(x_b,t,u)&=&(\wt d+\wt s)(x_b)\,\frac{d\hat\sigma_1}{dt}+\frac{1}{2}\,\wt u (x_b)\left[\frac{d\hat\sigma_2}{dt}+\frac{d\hat\sigma_4}{dt}-\frac{d\hat\sigma_4}{du}\right]+g(x_b)\,\frac{d\hat\sigma_7}{dt}\\
L_d(x_b,t,u)&=&(\wt u+\wt s)(x_b)\,\frac{d\hat\sigma_1}{dt}+\frac{1}{2}\,\wt d (x_b)\left[\frac{d\hat\sigma_2}{dt}+\frac{d\hat\sigma_4}{dt}-\frac{d\hat\sigma_4}{du}\right]+g(x_b)\,\frac{d\hat\sigma_7}{dt}\\
L_s(x_b,t,u)&=&(\wt u+\wt d)(x_b)\,\frac{d\hat\sigma_1}{dt}+\frac{1}{2}\,\wt s (x_b)\left[\frac{d\hat\sigma_2}{dt}+\frac{d\hat\sigma_4}{dt}-\frac{d\hat\sigma_4}{du}\right]+g(x_b)\,\frac{d\hat\sigma_7}{dt}
\ee
and for example
\be
L_u(x_a,u,t)=(\wt d+\wt s)(x_a)\,\frac{d\hat\sigma_1}{du}+
\frac{1}{2}\,\wt u(x_a)\left[\frac{d\hat\sigma_2}{dt}+\frac{d\hat\sigma_4}{du}-
\frac{d\hat\sigma_4}{dt}\right]+g(x_a)\,\frac{d\hat\sigma_7}{du}.
\label{diff1}
\ee
In the last formula  we have used  the fact that $\hat\sigma_2$ is symmetric under $t\leftrightarrow u$.
Here $u_V$ and $d_V$ are the usual valence quark PDFs:
\be
u_V=u-\bar u,\quad d_V=d-\bar d
\label{udv}
\ee
and we have used the notation:
\be
s_V=s-\bar s,\qquad \widetilde q=q+\bar q\,.
\label{tilda}
\ee

Using the strong bound on $(s - \bar s)$
obtained from neutrino experiments $s_V \leq 0.025$ \cite{Bourrely:2007if}
one may safely neglect the contribution of $s_VD_{s_V}^h$ and we have:
\be
&&E^h\frac{d\,\sigma_{pp}^{h^+-h^-}}{d^3P^h} =
\frac{1}{\pi}\int dx_a\,dx_b\,\frac{1}{z }\sum_{q=u,d}D_{q_V}^{h^+}(z )
\big( q_V(x_a)\,L_q(x_b,t,u)+q_V(x_b)L_q(x_a,u,t)\big)\label{shs}
\ee
This implies that  the contribution of
$(s-\bar s)D_s^{h^+-h^-}$ is expected to be within the experimental error and negligible, also
the large
uncertainties in $D_s^h$ should not affect the results
for $\sigma_{pp}^{h^+-h^-}$.

Note that in LO only 4 partonic cross sections contribute to the difference
$d\sigma_{pp}^{h^+-h^-}$:

-- scattering of flavour-unlike quarks $\hat\sigma_1$,

-- scattering of flavor-like quarks $\hat\sigma_2$ and $\hat\sigma_4$ and

-- quark-gluon scattering $ \hat\sigma_7$.

In NLO, the expressions for the difference cross sections are much more
complicated as all radiative corrections $\mathcal{O}(\alpha_s^3)$
must be accounted for. Nevertheless, again, one obtains only nonsinglet
contributions proportional to $D_{q_V}^{h^+}(z,Q^2)$.

\subsection{The difference cross section for $\pi^+-\pi^-$}

Eq. (\ref{sh}) considerably simplifies for pions if one assumed SU(2) isospin
symmetry, as used in all present analyses. It implies:
\be
D_u^{\pi^+-\pi^-} = -D_d^{\pi^+-\pi^-},\qquad D_s^{\pi^+-\pi^-}=0\cdot
\ee
Then $\sigma_{pp}^{\pi^+-\pi^-}$ is expressed solely in terms of $D_{u_V}^{\pi^+}$, enhanced by the best known valence-quark $u_V$:
\be
E^\pi\frac{d\,\sigma_{pp}^{\pi^+-\pi^-}}{d^3P^\pi} &=&
\frac{1}{\pi}\int dx_a\,dx_b\,\frac{1}{z }\,D_{u_V}^{\pi^+}(z )\times\nn
&&\times\left[ u_V(x_a)\,L_u(x_b,t,u)-d_V(x_a)\,L_d(x_b,t,u)+u_V(x_b)L_u(x_a,u,t)
-d_V(x_b)L_d(x_a,u,t)\right]\label{Kp}
\ee

\subsection{The difference cross section for $K^+-K^-$}

The  formula for $\sigma_{pp}^{K^+-K^-}$ strongly simplifies if we assume:
\be
 D_d^{K^+-K^-}=0\label{Kd}
\ee
This seems a reasonable physical  assumption that follows from the quark content of
$K^\pm = (\bar s ,u)$, $(s,\bar u)$ and is used in all current analyses.
Then $\sigma_{pp}^{K^+-K^-}$ depends solely on one non-singlet combination of FFs
$D_{u_V}^{K^+}$, multiplied by the large valence $u_V$-quark distributions:
\be
E^K\frac{d\,\sigma_{pp}^{K^+-K^-}}{d^3P^K} &=&
\frac{1}{\pi}\int dx_a\,dx_b\,\frac{1}{z }\,D_{u_V}^{K^+}(z )
\big[ u_V(x_a)\,L_u(x_b,t,u)+u_V(x_b)L_u(x_a,u,t)\big]\label{Kpm}
\ee

This expression may be used as a test for the assumption (\ref{Kd}).

\subsection{The difference cross section for  $K^\pm$ and $K_s^0$ }\label{K00}

 If in addition to the charged  $K^\pm$ also  neutral kaons $K_s^0=(K^0+\bar K^0)/\sqrt 2$
   are measured,  no new FFs are introduced into the cross-sections.
This is a consequence of SU(2) isospin symmetry, according to which
$(K^+, K^0)$ and $(K^-, \bar K^0)$ form isospin doublets, and we have:
\be
&&D_u^{K^++K^-}(z,Q^2)=D_d^{K^0+\bar K^0}(z,Q^2)=2D_d^{K_s^0}(z,Q^2)\nn
&&D_d^{K^++K^-}(z,Q^2)=D_u^{K^0+\bar K^0}(z,Q^2)=2D_u^{K_s^0}(z,Q^2)\nn
&&D_s^{K^++K^-}(z,Q^2)=D_s^{K^0+\bar K^0}(z,Q^2)=2D_s^{K_s^0}(z,Q^2).
\label{SU2K}
\ee
Note that the first relation involves favoured and unfavoured FFs,
while the second -- only unfavoured FFs. As explained in \cite{Albino:2008fy},
due to the nature of the DGLAP evolution, these constraints are independent of $Q^2$, as
constraints that follow from symmetry should be.

We write Eqs. (\ref{SU2K}) in the form:
\be
D_u^{K^+ + K^--2K_s^0}=-D_d^{K^+ + K^--2K_s^0}={(D_u-D_d)}^{K^+ + K^-} \label{SU2kaon}\ee
\be
D_s^{K^+ + K^--2K_s^0}= D_c^{K^+ + K^--2K_s^0}=D_b^{K^+ + K^--2K_s^0}=D_g^{K^+ + K^--2K_s^0}=0.
\label{SU2kaons}
\ee
Then, as   shown in \cite{Christova:2008te}, the combination
\be
 \sigma ^{K^+}+\sigma ^{K^-}-2\sigma ^{K_s^0}
\ee
 in the three types of inclusive processes, $K=K^\pm, K_s^0$:
 \be
&&e^++e^-\to K+X,\label{e+e-Ks}\\
&&e+N\to e +K+X,\qquad N=p,d, \label{SIDISKs}\\
&&p+p\to K+X,\label{ppKs}
 \ee
measure the same NS combination of FFs, namely
$(D_u-D_d)^{K^++K^-}$.  This result relies  only on SU(2) invariance for the kaons and
does not involve {\it any} assumptions about PDFs or  FFs; it holds in any order in QCD.

For inclusive hadron production,  from Eqs.~(\ref{Ks}) and (\ref{SU2kaon}) we obtain:
\be
\hspace*{-1cm}E^K\frac{d\s_{pp}^{ K^++K^--2K_S^0}}{d^3P^K}&&=\frac{1}{\pi}
\int\, dx_a\int\, dx_b\, \frac{1}{z }\,\times\nn
&&\hspace*{-1.5cm}\times\Big\{\left[ \tilde u(x_a)[\tilde d(x_b) + \tilde s(x_b)] -
\tilde d(x_a)[\tilde u(x_b) + \tilde s(x_b)]\right]\frac{d\hat\sigma_1}{dt}+\nn
&&\hspace*{-1.5cm}+\frac{1}{2}\,\left[u(x_a)u(x_b)+\bar u(x_a)\bar u(x_b) -
[d(x_a)d(x_b)+\bar d(x_a)\bar d(x_b)]\right]\frac{d\hat\sigma_2}{dt}+\nn
&&\hspace*{-1.5cm}+\left[d(x_a)\bar d(x_b)-u(x_a)\bar u(x_b)\right]
\left[\frac{d\hat\sigma_3}{dt}-\frac{d\hat\sigma_4}{dt}\right]+[\tilde u(x_a)-\tilde d(x_a)]g(x_b)\,\frac{d\hat\sigma_7}{dt}\nn
&&\hspace*{-1.5cm}
+\left[(x_a\leftrightarrow x_b),(t\leftrightarrow u)\right]\Big\}\,D_{u-d}^{K^++K^-}(z ).\label{K0}
\ee
Note that $D_q^{h +\bar h}$ can be determined directly from $e^+e^- \to h+X$ and thus without assumptions,
which implies  that  $D_q^{h +\bar h}$ should be roughly the same for all sets of FFs.
If this appears the case, then use of Eq.~(\ref{K0}) to compare the STAR data
on $K^+ + K^- -2K_s^0$ can be used as a test for the consistency of the data
with SU(2) invariance.


\section{Comparison to STAR data }

In this section we study numerically the impact of the different sets of FFs on the $p_T$-spectra
of the final hadrons in inclusive $pp$-collisions and compare it to STAR data.

Based on the theoretical NLO pQCD-formulas, we examine the sensitivity of the $p_T$-spectra to the FFs.
We use STAR data from 2006, 2007 and 2012 to try to distinguish among the different FFs.
First, we compare the cross sections for $d\s_{pp}^{h^{\pm}}$ and $d\s_{pp}^{h^+-h^-}$,
calculated with different sets for $D_{q}^h$, to the measured hadron yield.
Second, we try to obtain a fit of the FFs and compare it to the available parametrizations.

Note that there is a principle difference between the commonly used formula for
$\pi^\pm$-production and the one for $\pi^+-\pi^-$. In $\pi^\pm$ production
FFs with all flavours $D_i^h,\,\, i=g,u,d,s,\bar u, \bar d,\bar s $ enter the cross section,
while in the formula for $\pi^+-\pi^-$ it is only the valence-quark FFs that enter -- the non-singlet
combinations $D_{u_V}^h$ and $D_{d_V}^h$. Note also that, being non-singlets, in their $Q^2$-evolution
they do not mix with other FFs and this property holds in all orders in QCD.

Finally we use the preferred sets of FFs obtained in the above studies to make predictions
for the $p_T$ spectra of the charged pions and kaons produced in inclusive $pp$ collisions
at the planned NICA accelerator.

The results  are presented separately for final pions and kaons.

We  consider the following sets of NLO in pQCD FFs for pions:
HKNS-07 \cite{Hirai:2007cx}, AKK-08 \cite{Albino:2008fy},
DSEHS-14~\cite{deFlorian:2014xna}, LSS-15 \cite{Leader:2015hna} and for kaons:
HKNS-07 \cite{Hirai:2007cx}, AKK-08 \cite{Albino:2008fy}, DSEHS-17~\cite{deFlorian:2017lwf}.
These sets of FFs differ by the different processes, whose data they use, and by the used assumptions.
Only AKK and DSEHS use the inclusive $pp$-process as a source of information about the FFs.


We apply NLO approach \cite{Aversa:1988vb} to STAR data from 2006 \cite{STAR:2006xud}, 2007 \cite{STAR:2006nmo}
and 2012 \cite{Agakishiev:2011dc}.
The kinematics of the used measurements of STAR Collaborations in terms of $\sqrt S$, $p_T$
and the rapidity $y$ are:\\
STAR-2006, $\pi^\pm$:
\be
\qquad\qquad\qquad \sqrt S=200\, {\rm GeV},\qquad
0.3<p_T<10\, {\rm GeV}/c,\qquad
\qquad\vert y\vert <0.5\label{2006}
\ee
STAR-2007
\be
K^\pm:\qquad\qquad \sqrt S=200\, {\rm GeV},\qquad
0.25<p_T<2.2\, {\rm GeV}/c,\qquad
\qquad\vert y\vert <0.5\label{2007a}
\ee
\be
K_s^0:\qquad\qquad \sqrt S=200\, {\rm GeV},\qquad
0.26<p_T<4.7\, {\rm GeV}/c,\qquad
\qquad\vert y\vert < 0.5\label{2007b}
\ee
STAR-2012, $\pi^\pm , K^\pm , K_s^0$:
\be
\qquad\qquad\qquad\sqrt S=200\, {\rm GeV},\qquad
3.0<p_T<15\, {\rm GeV}/c,\qquad
\qquad\vert y\vert < 0.5\label{2012}
\ee

The STAR Collaborations present their results for the hadronic yield $d^2 N/dp_T\,dy$,
extracted from the measured non-single diffractive (NSD) $p+p$ cross section, while our formulae
are for the differential cross sections $d\s^h /d^3P_h$. The relation between them reads:
\be
\s_{NSD}\,\frac{d^2N}{(2\pi)\,p_T\,dp_T\,dy}=\,E_h\,\frac{d\s^h_{pp}}{d^3P_h}(p_T, \theta (y))
\label{nsd}
\ee
where $\s_{NSD}$ is the measured total NSD $p+p$ cross section ($\s_{NSD}=30.0\pm 3.5$ mb)
and $E_h\,d\s^h_{pp}/(d^3P_h)$ is our formula for the corresponding cross section
--  Eqs.~(\ref{Ks}),  (\ref{Kp}), (\ref{Kpm}) or (\ref{K0}).

In high energy hadron collider physics for the particle with the negligible mass the
rapidity $y$ coincides with the pseudo-rapidity $\eta$:
\be
y\approx\eta\equiv
\frac{1}{2}\ln\frac{1+\cos\theta}{1-\cos\theta}=-\ln\tan\frac{\theta}{2}\, ,
\label{rapid}
\ee
and hence
\be
\theta = 2\arctan (e^{-\eta})\approx 2\arctan (e^{-y})\, .
\label{theta}
\ee
As data from the STAR collaboration is presented averaged over rapidity:
$\vert y\vert <0.5$, which corresponds to $\theta \approx [\pi/3;\pi/2]$,
we compare data to the following averaged over $y$ expressions:
\be
\left.\s_{NSD}\,\frac{d^2N}{(2\pi)\,p_T\,dp_T\,dy}\right\vert_{\vert y\vert <0.5}=
\,\int_{-0.5}^{0.5}dy \,E_h\,\frac{d\s^h_{pp}}{d^3P_h}(p_T, \theta (y))\, .
\label{rap-12}
\ee

\subsection{Results for charged pions}

We calculate the $p_T$ spectra of $\pi^\pm$ using the parametrizations
for $D_i^{\pi^\pm}$ obtained in HKNS-07, AKK-08, DSEHS-14 and LSS-15, and compare it to data
on $\pi^{\pm}$-spectra measured by the STAR Collaborations STAR-2006 and STAR-2012 at
$\sqrt S = 200$ GeV for different ranges of $p_T$
and rapidity $\vert y\vert <0.5$ -- see Eqs.~(\ref{2006}) and (\ref{2012}).
All parametrizations are obtained within the same theoretical framework -- NLO in QCD, assuming
that isospin SU(2) symmetry of the $u$ an $d$ quarks for the favored and
unfavored fragmentation functions holds at the initial scale $Q^2_0$:
\be
D_u^{\pi^+}=D_{\bar d}^{\pi^+},\qquad D_{\bar u}^{\pi^+}=D_{ d}^{\pi^+}=D_s^{\pi^+}=D_{\bar s}^{\pi^+}\cdot
\ee
Using charge conjugation invariance of strong interactions we have, in  total, 3 independent FFs:
\be
D_u^{\pi^+},\quad  D_d^{\pi^+}, \quad D_g^{\pi^+}\cdot
\ee

The obtained FFs differ by the used data and the used parametrizations at $Q^2_0$.
Bellow we give the main sources of data of the used here FFs.

$\bullet$ In HKNS-07 the parametrizations and their uncertainties are obtained both in LO and NLO from
analysis of the data on $e^+e^-\to \pi^\pm +X$. We use the NLO parameters and show the error band based
on the Hessian method \cite{Pumplin:2001ct} and the error matrix provided by HKNS
{\cite{Hirai:2007cx}. The analysis for kaons proceeds in the same manner.

$\bullet$ In AKK-08 the parametrizations are obtained  from  analysis of the data on
$e^+e^-\to \pi^\pm +X$ and $pp\to \pi^\pm +X$.

$\bullet$ In DSEHS-14 the parametrizations are obtained in a global  analysis of the data on
$e^+e^-\to \pi^\pm +X$,  $l+N\to \pi^\pm +X$ and $pp\to \pi^\pm +X$
including the STAR-2012 data. The uncertainties estimated within the Hessian approach
are presented as shaded areas in their figures and the error matrix can be provided upon request.
In turn, in our analysis, both for pions and kaons, we estimate the scale uncertainty when $Q$
varies in the range $p_T/2\leqslant Q \leqslant 2p_T$.

$\bullet$ In LSS-15 the parametrizations are obtained using the published HERMES data from 2013
on pion multiplicities from SIDIS $l+N\to \pi^\pm +X$ on proton and deutron targets.

The results of our comparison to pion production are shown in Figs.~\ref{figu1}--\ref{figu5}
and in Table~\ref{tab1}. All $p_T$ spectra for pions (and kaons in the next
section) at the high energy $\sqrt{S}=200\,{\rm GeV}$ are presented for
$p_T\geqslant 1\, {\rm GeV}/c$ suitable for the perturbative QCD approach.
This is consistent with the constraint on the $z_{min} = 0.01$, and
${\rm min}\,p_T = z_{min}\sqrt{S}/2$.

\begin{figure}[h!]
\centering
\includegraphics[width=0.45\textwidth]{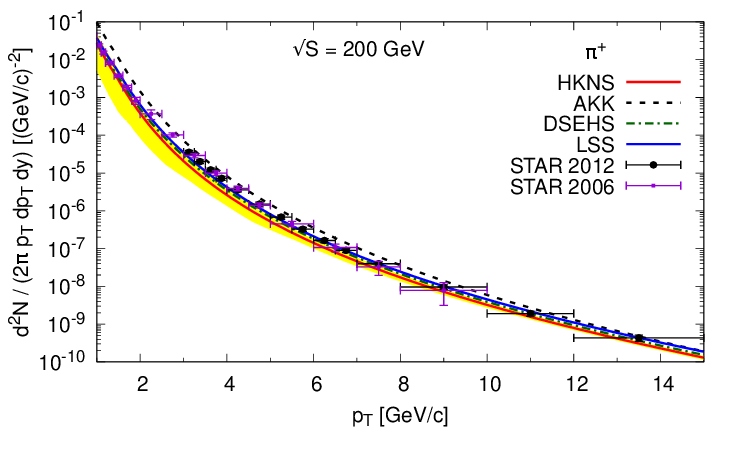}
\hfill
\includegraphics[width=0.45\textwidth]{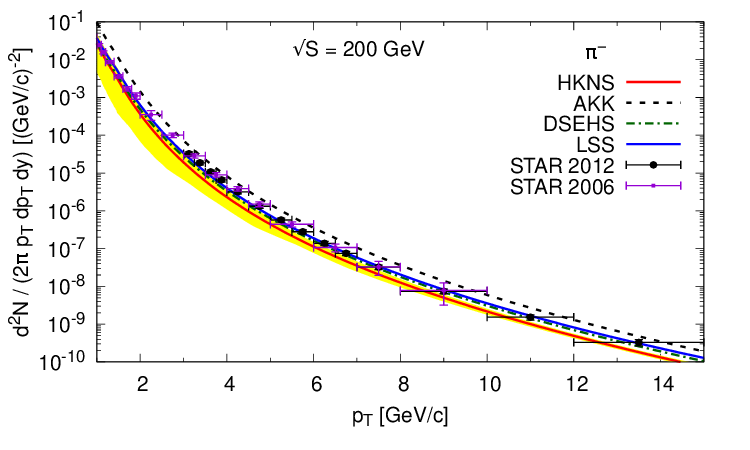}
\caption{The cross sections $\s_{pp}^{\pi^+}$ (left) and $\s_{pp}^{\pi^-}$
(right) calculated for different sets of FFs, compared to STAR-2006
and STAR-2012 data.
The one-$\sigma$ uncertainty band for HKNS FFs estimated by the
Hessian method is shown.}
\label{figu1}
\end{figure}
\begin{figure}[h!]
\centering
\includegraphics[width=0.45\textwidth]{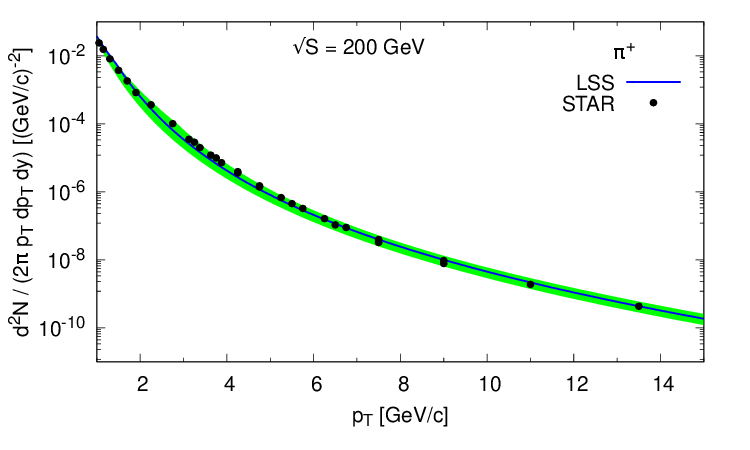}
\hfill
\includegraphics[width=0.45\textwidth]{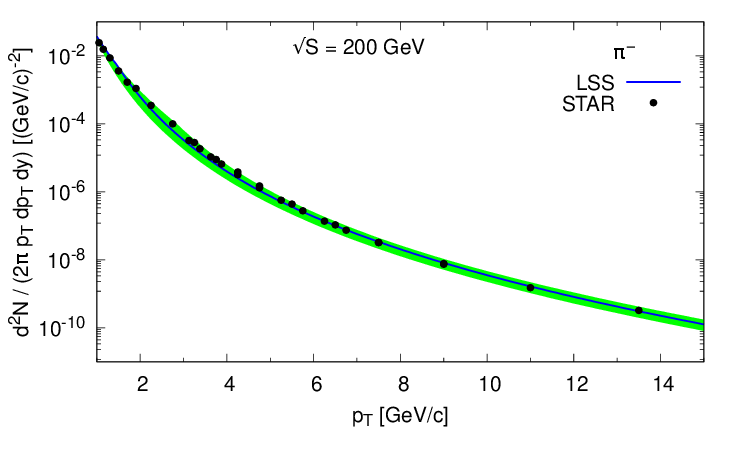}
\caption{The best fit LSS together with $\s_{pp}^{\pi^+}$ (left) and $\s_{pp}^{\pi^-}$
(right) STAR data. The theoretical uncertainties bands indicate the scale
$Q$ variation in the range $p_T/2\leqslant Q \leqslant 2p_T$.}
\label{figu2}
\end{figure}
\begin{figure}[h!]
\centering
\includegraphics[width=0.45\textwidth]{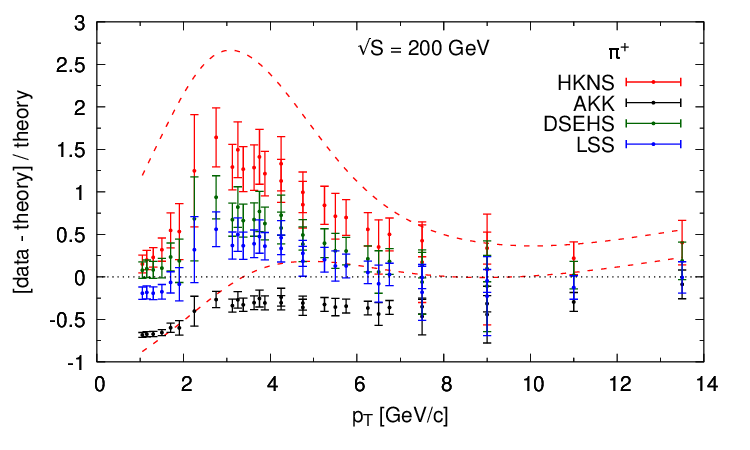}
\hfill
\includegraphics[width=0.45\textwidth]{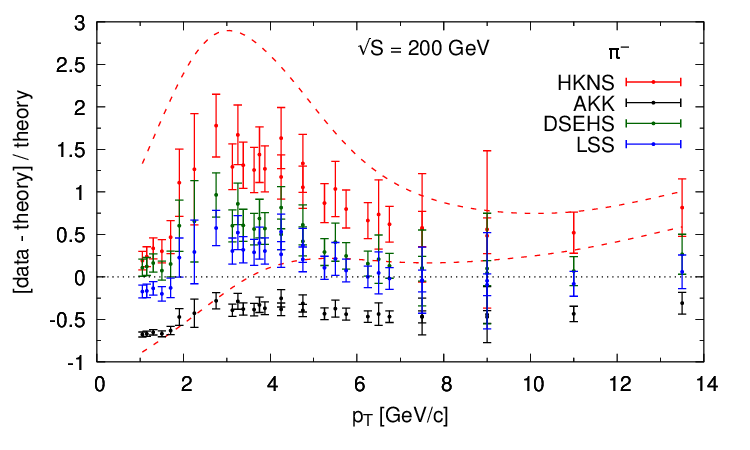}
\caption{The ratio (data-theory)/theory for the cross sections:
$\s_{pp}^{\pi^+}$ (left) and $\s_{pp}^{\pi^-}$ (right) for different FFs.
The experimental error bars and theoretical uncertainties for HKNS 
(dashed) are shown.}
\label{figu3}
\end{figure}

In Figs.~\ref{figu1}--\ref{figu3}, we show the agreement of the
$p_T$-spectra for the charged pions $\s_{pp}^{\pi^{\pm}}$ calculated
for the different sets of FFs with the STAR-2006 and STAR-2012 data.

\begin{table}[h!]
\caption
{Values of $\chi^2$, Eq.~(\ref{chi_def}), calculated for the single and
difference cross sections in the pion production obtained for
different FFs parametrizations.}
\begin{tabular}{*4c}
\hline\hline
~~~FFs~~~ &   $~~~\pi^+~~~$ & $~~~\pi^-~~~$ & $~~\pi^+ - \pi^-~~$ \\
\hline
HKNS  & 10.2  & 11.7  & 0.136 \\
AKK   & 77.2  & 83.3  &   -   \\
DSEHS & 4.55  & 4.39  & 0.148 \\
LSS   & 2.83  & 2.46  & 0.143 \\
\hline\hline
\end{tabular}
\label{tab1}
\end{table}

Using the values of $\chi^2$ defined as
\be
\chi^2=\frac{1}{N}\sum_i^N\frac{(\s_i^{exp}-\s_i^{th})^2}{\Delta_i^2},
\label{chi_def}
\ee
where $\Delta_i^2$ incorporates both statistical and systematic
experimental errors, we are able to determine the best fit to the data.
In our analysis for the pions, we took into account the combined STAR data
for $p_T>1\, {\rm GeV}^2$ with $N=30$ experimental points (STAR 2006 and STAR 2012).
We find that the NLO LSS results agree the best with the data and
the DSEHS fit which is partially based on the STAR-2012 data is comparably
good as LSS. In Table~\ref{tab1}, we collect the values of $\chi^2$,
Eq.~(\ref{chi_def}), for the single charged pions $\pi^{\pm}$ and also for the
difference $\pi^+ - \pi^-$ obtained with the use of HKNS, AKK, DSEHS and LSS
sets of FFs.

\begin{figure}[h!]
\centering
\includegraphics[width=0.45\textwidth]{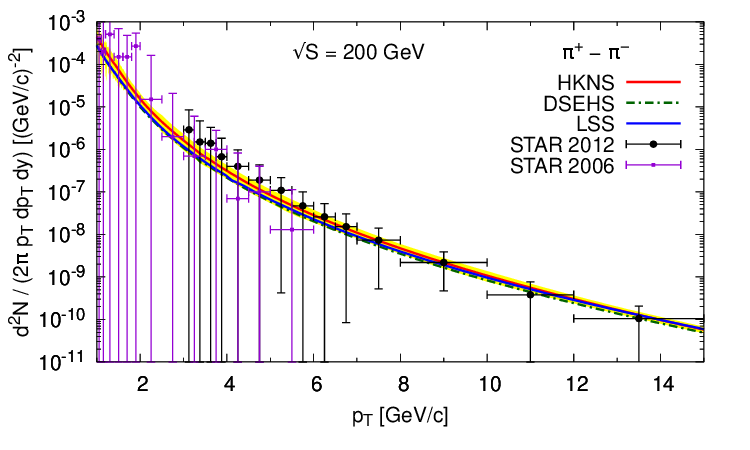}
\hfill
\includegraphics[width=0.45\textwidth]{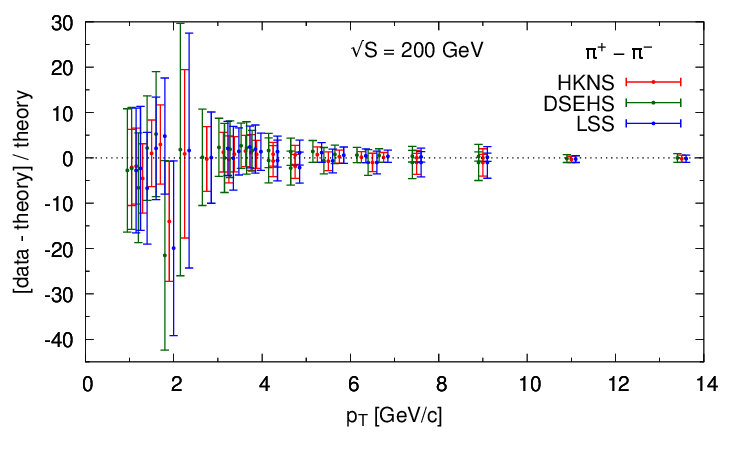}
\caption{Left: The difference cross section $\s_{pp}^{\pi^+-\pi^-}$ calculated
for different sets of FFs, compared to the STAR data.
The one-$\sigma$ uncertainty band for HKNS FFs is shown.
Right: The ratio (data-theory)/theory for the difference cross section.
The experimental error bars are shown. For better visibility points for
DSEHS (LSS) are displaced by -0.1 (+0.1) in $p_T$.}
\label{figu4}
\end{figure}

Fig.~\ref{figu4} shows a comparison of the difference cross section
$\s_{pp}^{\pi^+-\pi^-}$ calculated for different sets of FFs with the
STAR-2006 and STAR-2012 data. Since the experimental errors are relatively large,
all FFs parametrizations are comparably good for this nonsinglet case.
Finally, in Fig.~\ref{figu5}, using LSS FFs, we compare NLO and LO
results for $\s_{pp}^{\pi+}$. The right panel shows the $K$ factor, defined as
\be
K = \frac{\left[ d\s_{pp}^{h}\right]^{\rm NLO}}{\left[d\s_{pp}^{h}\right]^{\rm LO}},
\label{K-factor}
\ee
where NLO calculations include the full $\alpha_s^3$ radiative NLO corrections
presented by Aversa et. al. in 1988, \cite{Aversa:1988vb}, and LO stands for
the LO Born terms. It is seen that NLO predictions provide a better match to the data
than LO ones, and the NLO corrections are more essential in the small $p_t$
range. The middle panel of Fig.~\ref{figu5} shows the scale dependence of
$\s_{pp}^{\pi^+}$ at LO and NLO where the scale $Q$ may vary from $p_T/2$
to $2p_T$. It is seen that the scale dependence is clearly smaller at NLO.
The same occurs for the polarized cross section for $\pi^0$ production,
\cite{Jager:2002xm}.
\begin{figure}[h!]
\centering
\includegraphics[width=0.32\textwidth]{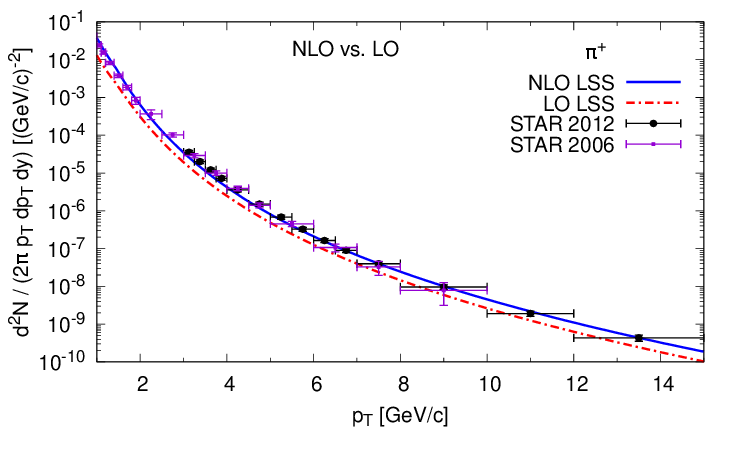}
\includegraphics[width=0.32\textwidth]{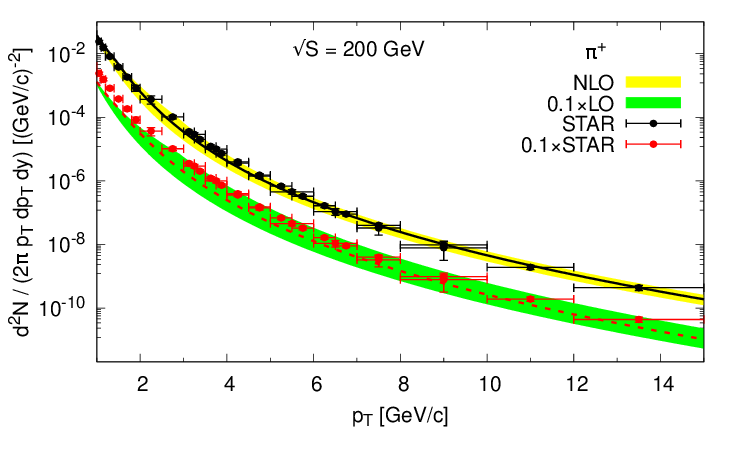}
\includegraphics[width=0.32\textwidth]{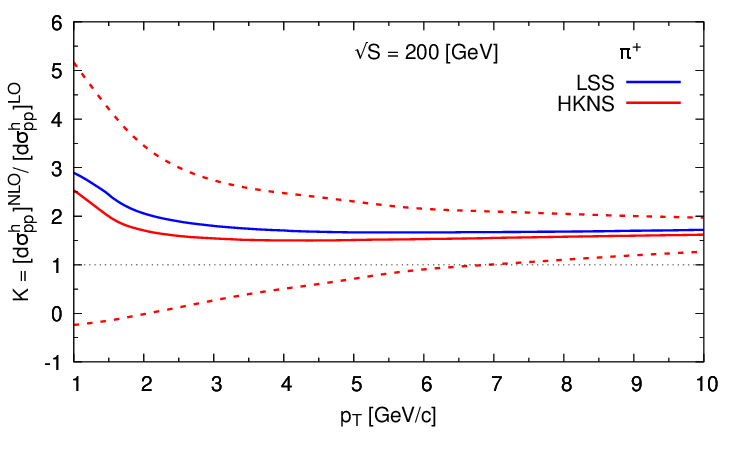}
\caption{Left: a comparison of the NLO and LO results for the cross section
$\s_{pp}^{\pi^+}$ with the use of LSS FFs together with the STAR data.
Middle: the scale dependence of $\s_{pp}^{\pi^+}$ at LO and NLO for LSS FFs..
The shaded bands indicate the uncertainties from varying the $Q$ scale
in the range $p_T/2\leqslant Q \leqslant 2p_T$. The solid (dashed) lines
are for the choice $Q=p_T$. For better visibility, the LO results
(and the STAR data, respectively) are rescaled by 0.1. 
Right: the $K$ factor, Eq.~(\ref{K-factor}), vs $p_T$ for the HKNS and LSS
results at the choice $Q=p_T$. The error band for HKNS is shown (dashed).}
\label{figu5}
\end{figure}

\subsection{Results for charged kaons}

We calculate the $p_T$  spectra of $K^+$ using the parametrizations
for $D_i^{K^+}$ obtained in HKNS-07, AKK-08 and DSEHS-17, and compare it to data
on $K^{\pm}$-spectra measured by the STAR Collaborations 
STAR-2007 and STAR-2012 at $\sqrt S = 200$ GeV for different ranges of $p_T$
and rapidity $-0.5<y<0.5$ -- see Eqs.~(\ref{2007a}) and (\ref{2012}).
All considered parametrizations of $D_q^{K^+}$ are obtained within the same theoretical framework -- 
NLO in QCD\footnote{In an interesting study~\cite{Demirci:2018uun}, Demirci and Ahmadov have
considered various models of FFs and also the effect of HT in charged kaon production and
suggested that HT effects may be large in certain kinematic regions.},
using the simplifying assumption that at the initial scale $Q^2_0$ all unfavored
fragmentation functions are equal:
\be
D_{\bar u}^{K^+}(z,Q_0^2)=D_{\bar d}^{K^+}(z,Q_0^2)=D_{d}^{K^+}(z,Q_0^2)=D_{s}^{K^+}(z,Q_0^2).\label{unf1}
\ee
If neutral kaons are considered then SU(2) invariance relates the FFs of
$K^0$ to those for charged kaons and thus no new FFs are introduced, and,
as shown, this leads to relation (\ref{K0}). If only charged kaons are
involved there are no SU(2) constraints. SU(2) invariance enters the game
only when $K^0$ are measured as well.
Taking into account also charge conjugation invariance of strong interactions
we end up, in  total, with  4 independent kaon FFs:
\be
D_u^{K^+}, \quad D_{\bar u}^{K^+}, \quad D_{\bar s}^{K^+},\quad D_g^{K^+}\cdot
\ee

Below we give the main characteristics of the FFs used here.

$\bullet$ In HKNS-07 the parametrizations are obtained both in LO and NLO in QCD from an analysis
of the data on $e^+e^-\to K^\pm +X$. We use the NLO parameters.

$\bullet$ In AKK-08 the parametrizations are obtained in NLO from an analysis of the data
on $e^+e^-\to K^\pm +X$ and $pp\to K^\pm +X$. No uncertainties of the obtained parametrizations
are presented.

$\bullet$ In DSEHS-17 the parametrizations are obtained in a global analysis
of the data on $e^+e^-\to K^\pm +X$,  $l+N\to K^\pm +X$ and $pp\to K^\pm +X$
including the STAR-2012 data. The parameters of the FFs are presented in NLO, using
the same assumption Eq.~(\ref{unf1}) that all unfavoured FFs at the initial scale $Q^2_0$
are the same, but here at NLO.

Our results for kaons are presented in Figs.~\ref{figu6}--\ref{figu10} and
in Table~\ref{tab2} in a similar way as those for pions.
\begin{figure}[h!]
\centering
\includegraphics[width=0.45\textwidth]{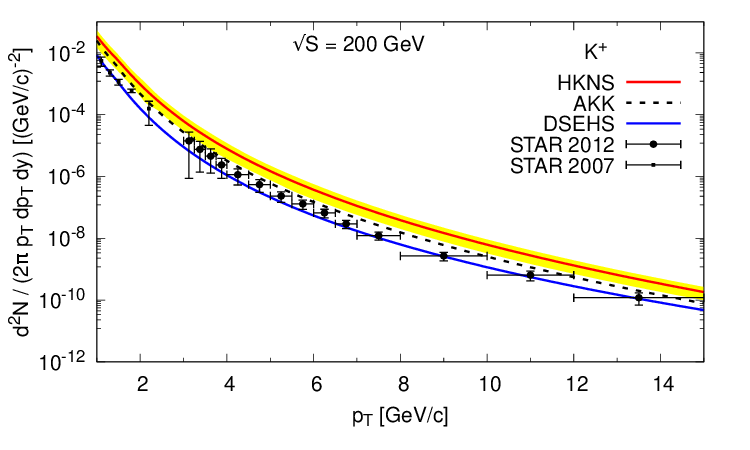}
\hfill
\includegraphics[width=0.45\textwidth]{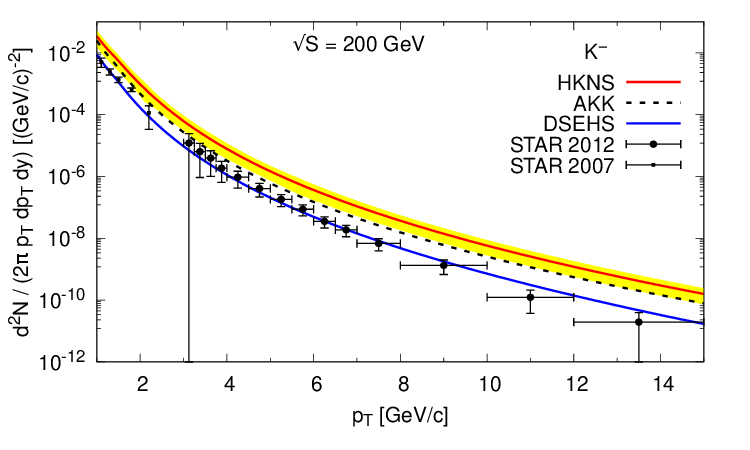}
\caption{The cross sections $\s_{pp}^{K^+}$ (left) and $\s_{pp}^{K^-}$ (right)
calculated for different sets of FFs, compared to STAR-2007 and STAR-2012 data.
The one-$\sigma$ uncertainty band for HKNS FFs estimated by the
Hessian method is shown.}
\label{figu6}
\end{figure}
\begin{figure}[h!]
\centering
\includegraphics[width=0.45\textwidth]{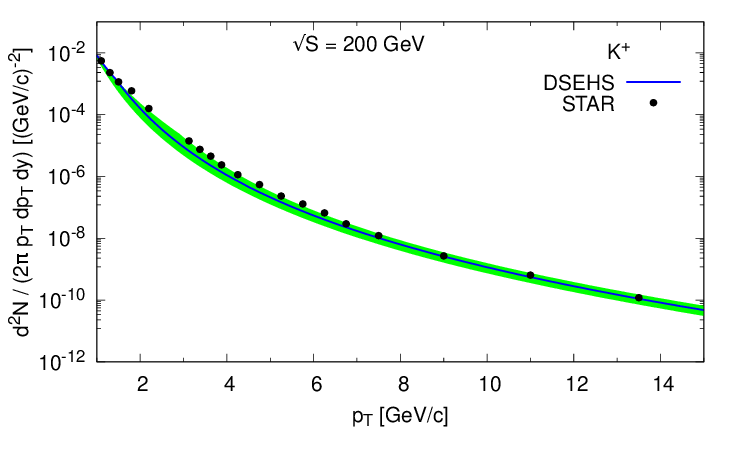}
\hfill
\includegraphics[width=0.45\textwidth]{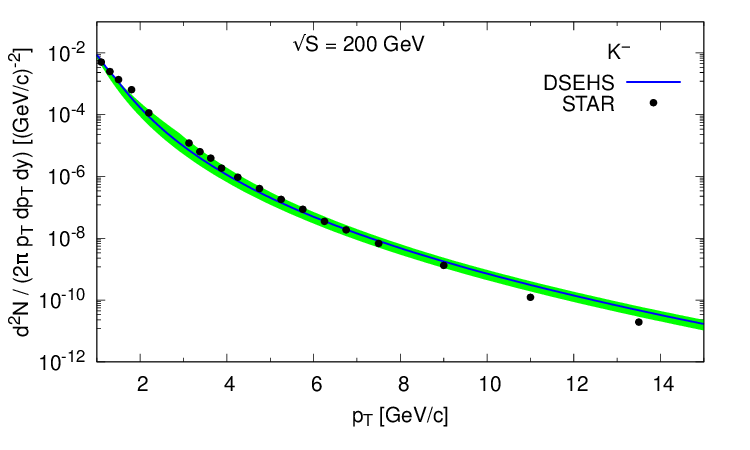}
\caption{The best fit DSEHS together with $\s_{pp}^{K^+}$ (left) and
$\s_{pp}^{K^-}$ (right) STAR data. The theoretical uncertainties bands
indicate the scale $Q^2$ variation in the range
$p_T/2\leqslant Q \leqslant 2p_T$.}
\label{figu7}
\end{figure}
\begin{figure}[h!]
\centering
\includegraphics[width=0.45\textwidth]{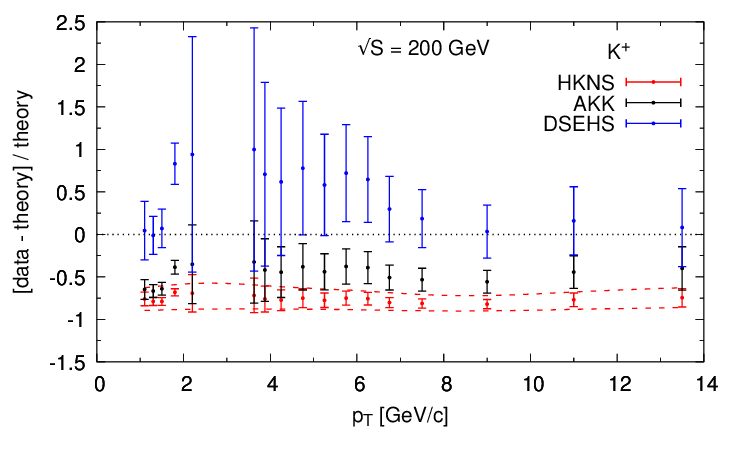}
\hfill
\includegraphics[width=0.45\textwidth]{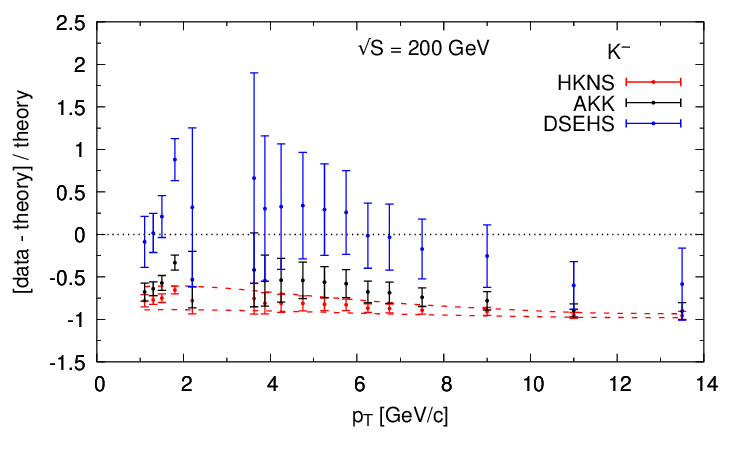}
\caption{The ratio (data-theory)/theory for the cross sections:
$\s_{pp}^{K^+}$ (left) and $\s_{pp}^{K^-}$ (right) for different FFs.
The experimental error bars and theoretical uncertainties for HKNS 
(dashed) are shown.}
\label{figu8}
\end{figure}
\begin{table}[h!]
\caption
{Values of $\chi^2$, Eq.~(\ref{chi_def}), calculated for the single and
difference cross sections in the kaon production obtained for
different FFs parametrizations.}
\begin{tabular}{*4c}
\hline\hline
~~~FFs~~~ &   $~~~K^+~~~$ & $~~~K^-~~~$ & $~~K^+ - K^-~~$ \\
\hline
HKNS  & 124   & 222   & 0.335 \\
AKK   & 16.0  & 34.1  &   -   \\
DSEHS & 1.18  & 1.31  & 0.329 \\
\hline\hline
\end{tabular}
\label{tab2}
\end{table}

From Figs.~\ref{figu6}--\ref{figu8} and also from the $\chi^2$
analysis, Eq.~(\ref{chi_def}), it is seen that the DSEHS-17 FFs
which are based also on the STAR-2012 data provide the best fit.
We took into account the combined STAR data for $p_T>1\, {\rm GeV}^2$
with $N=17$ experimental points (STAR 2007 and STAR 2012).
In Table~\ref{tab2}, we collect the values of $\chi^2$,
Eq.~(\ref{chi_def}), for the single charged kaons $K^{\pm}$ and also
for the difference $K^+ - K^-$ obtained with the use of HKNS, AKK and
DSEHS sets of FFs.

\begin{figure}[h!]
\centering
\includegraphics[width=0.32\textwidth]{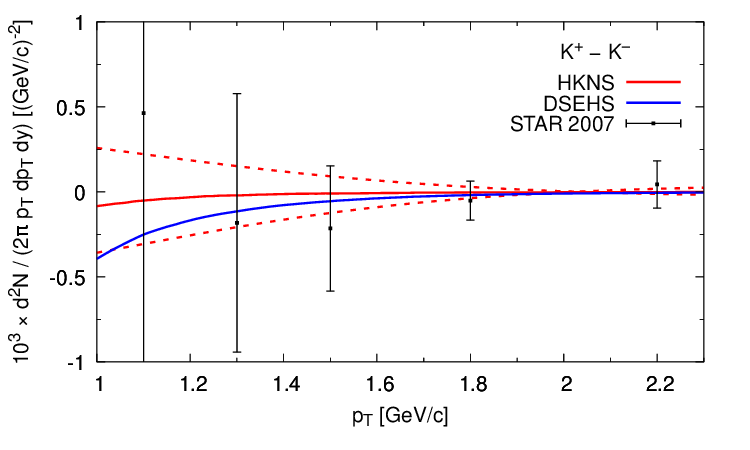}
\includegraphics[width=0.32\textwidth]{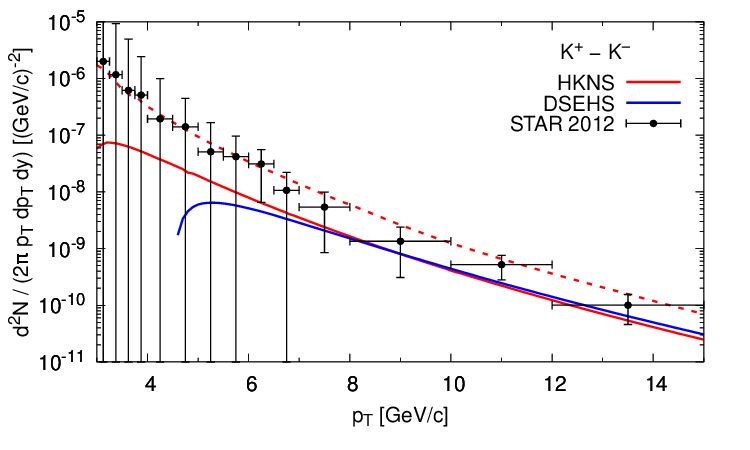}
\includegraphics[width=0.32\textwidth]{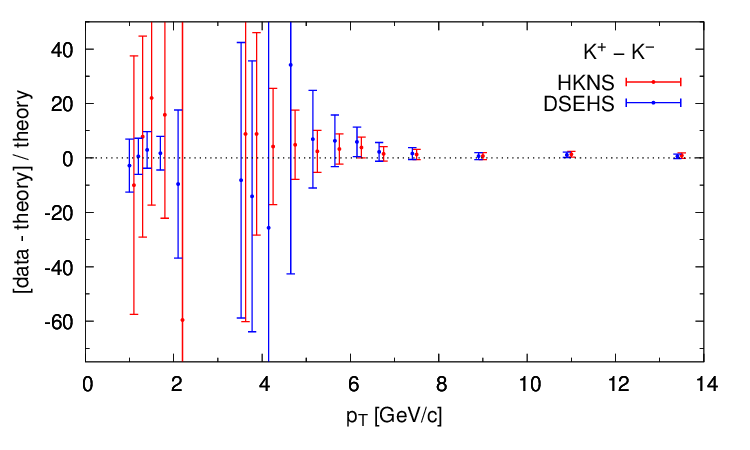}
\caption{The difference cross section $\s_{pp}^{K^+-K^-}$ calculated
for different sets of FFs, compared to the STAR data.
Left: for the small $p_T$ range. The one-$\sigma$ uncertainty band for
HKNS FFs is shown (dashed).
Middle: for $p_T>3\,{\rm GeV}/c$. Since the theoretical error for HKNS is
larger than the value of $\s_{pp}^{K^+-K^-}$ itself, we show only the upper
line of the uncertainty (dashed). 
Right: The ratio (data-theory)/theory for the difference cross section.
The experimental error bars are shown. For better visibility points for
DSEHS are displaced by -0.1 in $p_T$.}
\label{figu9}
\end{figure}

A comparison of the difference cross section $\s_{pp}^{K^+-K^-}$ obtained for different
sets of FFs with the STAR-2007 and STAR-2012 data is presented in Fig.~\ref{figu9}.
For some experimental points of STAR-2007 and also for the theoretical predictions
at $p_T<3\,{\rm GeV}/c$ for HKNS and $p_T<4.5\,{\rm GeV}/c$ for DSEHS $\s_{pp}^{K^+-K^-}<0$.
Again, similarly as for pions, the large experimental errors make useless the idea to
find the best fit.

A comparison of the NLO and LO predictions to the data for $\s_{pp}^{K+}$
presented in Fig.~\ref{figu10} shows, similarly as in the pion case, better
agreement of the NLO results with the data. From the middle panel one can
see the smaller scale dependence at NLO. The $K$ factor, Eq.~(\ref{K-factor}),
obtained for $Q=p_T$, presented in the right panel of Fig.~\ref{figu10},
varies between about 2 for $p_T=10\,{\rm GeV}/c$ to 3 for $p_T=1\,{\rm GeV}/c$.
\begin{figure}[h!]
\centering
\includegraphics[width=0.32\textwidth]{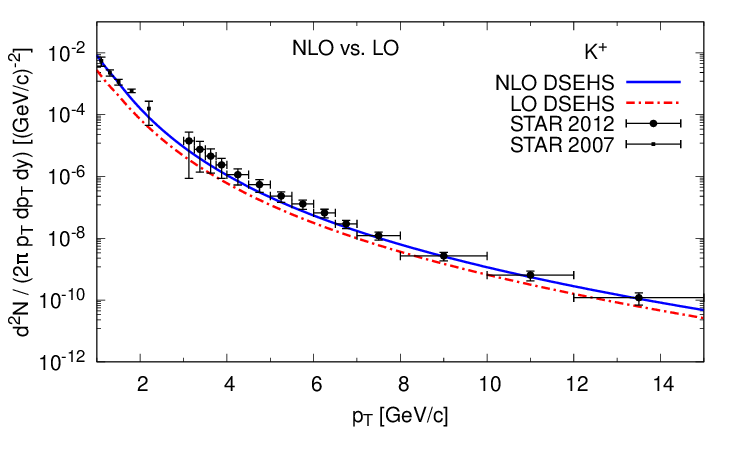}
\includegraphics[width=0.32\textwidth]{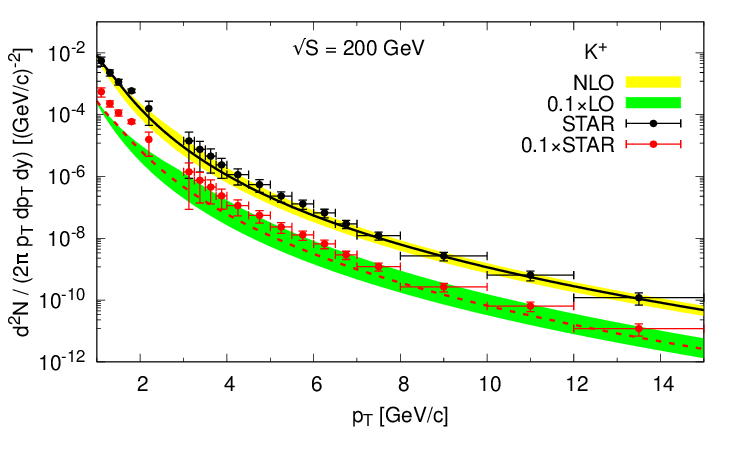}
\includegraphics[width=0.32\textwidth]{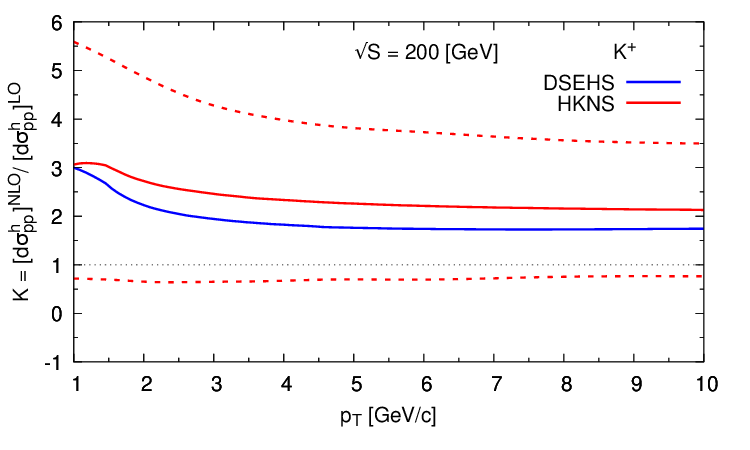}
\caption{Left: a comparison of the NLO and LO results for the cross section
$\s_{pp}^{K^+}$ with the use of DSEHS FFs together with the STAR data.
Middle: the scale dependence of $\s_{pp}^{K^+}$ at LO and NLO for DSEHS FFs.
The shaded bands indicate the uncertainties from varying the $Q$ scale
in the range $p_T/2\leqslant Q \leqslant 2p_T$. The solid (dashed) lines
are for the choice $Q=p_T$. For better visibility, the LO results
(and the STAR data, respectively) are rescaled by 0.1. 
Right: the $K$ factor, Eq.~(\ref{K-factor}), vs $p_T$ for the HKNS and DSEHS
results at the choice $Q=p_T$. The error band for HKNS is shown (dashed).}
\label{figu10}
\end{figure}

\subsection{Results for neutral and charged kaons $K^++K^--2K_s^0$}

It is important to recall that if not only charged, but neutral kaons are measured, as well,
and if SU(2) isospin symmetry holds no new kaon FFs are introduced. In this section, we shall
test SU(2) invariance of the $u$ ad $d$ quarks for kaons.

If the FFs of both charged and neutral kaons are extracted {\textit independently} from the data,
then they should satisfy Eq.~(\ref{SU2K}).
Only in AKK~\cite{Albino:2008fy} the neutral and charged
kaons FFs are obtained separately, without imposing SU(2)-invariance.
This allows to test Eq.~(\ref{SU2K}) and study the SU(2)- symmetry directly.
\begin{figure}[h!]
\centering
\includegraphics[width=0.32\textwidth]{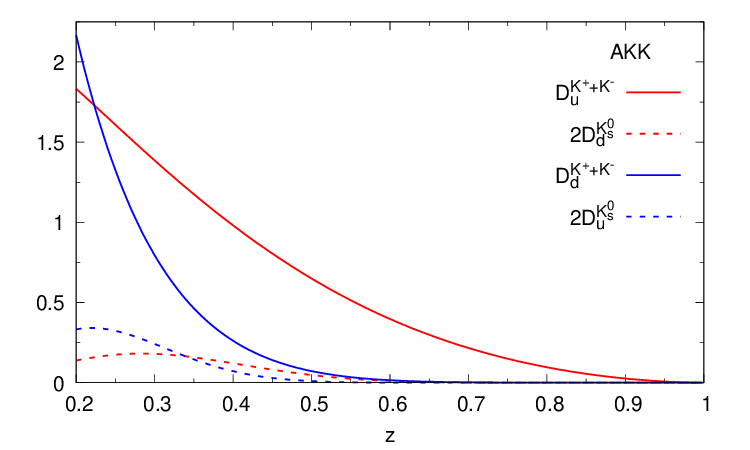}
\includegraphics[width=0.32\textwidth]{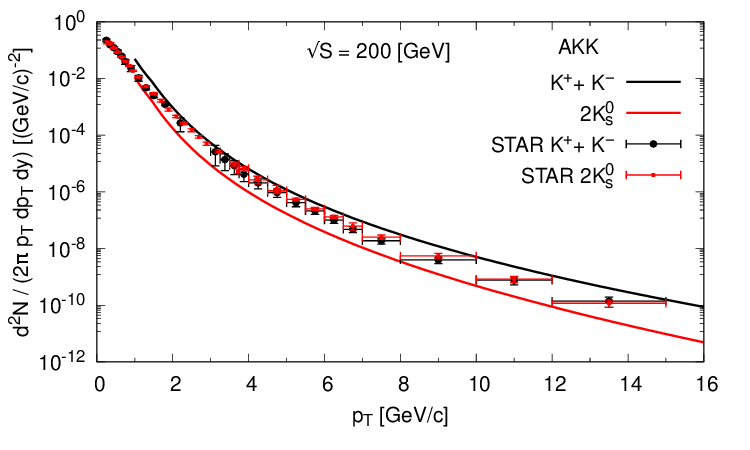}
\includegraphics[width=0.32\textwidth]{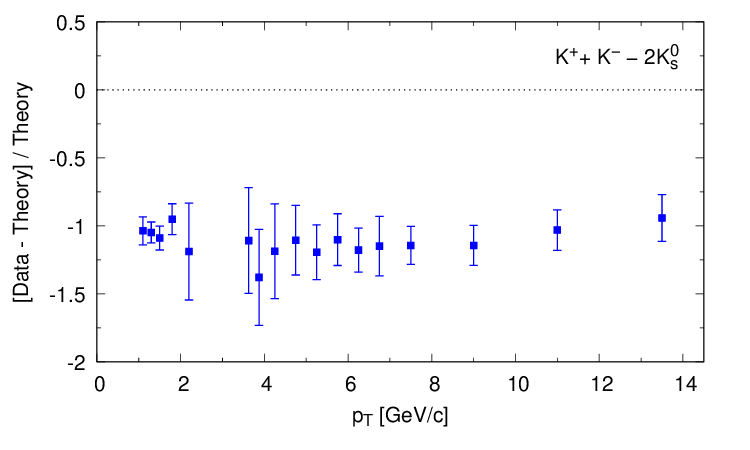}
\caption{Left: Test of the SU(2) invariance via Eq.~(\ref{SU2K}) for AKK FFs.
Middle: A comparison of the AKK predictions with the STAR data for $K^++K^-$ and $2K_s^0$.
Right: [Data - Theory]/Theory for $K^++K^--2K_s^0$. The experimental error bars
are shown.}
\label{figu11}
\end{figure}

In the left panel of Fig.~\ref{figu11}, we compare $D_u^{K^++K^-}$ with $2D_d^{K_s^0}$
and $D_d^{K^++K^-}$ with $2D_u^{K_s^0}$.
This clearly shows that isospin symmetry is badly broken for the AKK-08 kaon FFs.
It is also visible in the middle and right panels of the same figure.
The experimental STAR data for kaons where $\s^{K^++K^-} = \s^{2K_s^0}$
confirm the SU(2)-invariance. The theoretical predictions for HKNS and DSEHS
based on FFs for which the SU(2)-invariance is assumed also provides,
contrary to AKK, $\s^{K^++K^-} - \s^{2K_s^0}=0$.

In summary, the above studies clearly show that one can distinguish two types of
parametrizations of the FFs that describe the $p_T$-spectra of the produced {\textit charged}
kaons in $pp$ collisions:
HKNS and DSEHS, for which SU(2)-invariance holds - that agrees with the data - 
and AKK for which it does not. For this reason, making the predictions
for NICA, we shall not longer consider AKK a reasonable fit for kaons.

\section{Results  for NICA Project -- pions and kaons}

NICA (Nuclotron-based Ion Collider fAcility) is a new accelerator complex
under construction at the Joint Institute for Nuclear Research in Dubna.
In anticipation of the NICA results we have produced
estimates of various semi-inclusive cross-sections based on the discussed
above FFs. These should be useful for planning experiments.

The accelerated beams in NICA by design will consist of particles ranging from
protons and light nuclei to fully stripped gold ions. Beam energies for $p+p$
collisions will span $\sqrt{S} = 12-27\,{\rm GeV}$ with luminosity
$L_{pp}\geqslant 10^{30}\, {\rm cm^{-2}s^{-1}}$, and
$\sqrt{S_{NN}} = 4-11\,{\rm GeV}$ with averaged luminosity
$L_{\rm Au+Au}=10^{27}\, {\rm cm^{-2}s^{-1}}$ for heavy-ion $Au+Au$ collisions
\cite{Kekelidze:2020hdy,Taranenko:2020vqn}.
The expected ranges of $p_T$ and pseudo-rapidity $\eta$, Eq.~(\ref{rapid}),
in NICA Multi Purpose Detector (MPD) for identification of pions and kaons are:
\be
0.1\, {\rm GeV}/c < p_T < 2\, {\rm GeV}/c\\
0.1\,\pi < \theta < 0.9\,\pi \,\,\to\,\, |\eta| < 2\, ,
\ee
providing almost the whole forward hemisphere in which the final hadrons will
be detected.

Below, we present the NLO pQCD predictions for NICA energies and $p_T$
based on the viable sets of FFs elaborated in the previous sections.
We find it interesting to compare them with the existing data from the
STAR experiment in the Beam Energy Scan (BES) program at the Relativistic
Heavy Ion Collider (RHIC) on the hadron production in Au$+$Au collisions
at the energies similar to those planned at NICA \cite{STAR:2017sal}.

We use the phenomenological result that spectral shapes
and relative particle yields are similar in $p+p$ and peripheral $A+A$
collisions where the nuclear effects are negligible
\cite{PHENIX:2001hpc,STAR:2003jwm}.
Thus, one can regard the nuclei as an incoherent superposition of
partons and approximate the A$+$A collisions as a sum of independent
nucleon-nucleon (N$+$N) collisions. In this approach, the hard process
inclusive yield in nuclear collisions is expected to scale as $N_{\rm coll}$,
the average number of inelastic N$+$N collisions
\cite{PHENIX:2001hpc,STAR:2003jwm}, and the nuclear modification factor is
defined as a ratio:
\be
R_{AA}(p_T) = \frac{d^2N^{AA}/dp_T\,dy}{N_{\rm coll}\,d^2N^{NN}/dp_T\,dy}.
\label{scal1}
\ee
In the absence of nuclear modifications to hard scattering, $R_{AA}=1$,
and we obtain the scaling relation between inclusive yields in $p+p$
and peripheral ($R_{AA}=1$) $A+A$ collisions:
\be
\frac{d^2N^{AA}}{dp_T\,dy} = N_{\rm coll}\,\frac{d^2N^{pp}}{dp_T\,dy}\,,
\label{scal2}
\ee
where
\be
\frac{d^2N^{pp}}{2\pi\,p_T\,dp_T\,dy} =
\frac{1}{\s^{pp}_{inel}}\,E_h\,\frac{d\s^h_{pp}}{d^3P_h}.
\label{scal3}
\ee

In the first step, we illustrate the scaling, Eqs.~(\ref{scal2}) and
(\ref{scal3}) at high energy $\sqrt{S}=200\,{\rm GeV}$ by comparing our
NLO pQCD results for $p_T$ spectra of $\pi^+$ in $p+p$ collisions with
the STAR data on the semi-inclusive hadron production in the most peripheral
Au$+$Au collisions \cite{STAR:2006uve}. We used the best fit, i.e., LSS
parametrization of FFs. 
For thus chosen the most peripheral events, nuclear corrections can be neglected,
and the scaling works, i.e., the pQCD results for $p+p$ scaled up, according to
Eqs.~(\ref{scal2}) and (\ref{scal3}), by $N_{\rm coll}/{\s^{pp}_{inel}}$,
where $N_{\rm coll}$ is provided with the experimenta data, give the results
for Au+Au (most peripheral).
This is shown in Fig.~\ref{nfig1} where we plot
Au$+$Au data \cite{STAR:2006uve} for the most peripheral collisions
(centrality $60-80\%$)\footnote{Details on centralities and $N_{\rm coll}$
values estimations using a Glauber model can be found in
\cite{STAR:2008med}.}
and also NSD $p+p$ data \cite{STAR:2006xud,STAR:2006nmo,Agakishiev:2011dc}.
\begin{figure}[h!]
\centering
\includegraphics[width=0.45\textwidth]{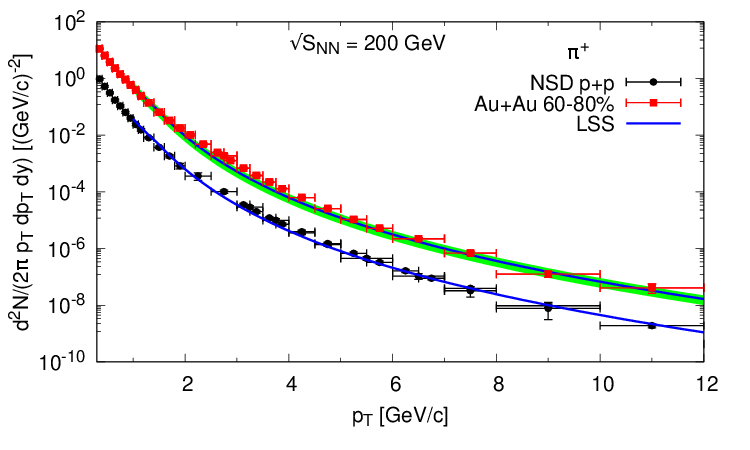}
\hfill
\includegraphics[width=0.45\textwidth]{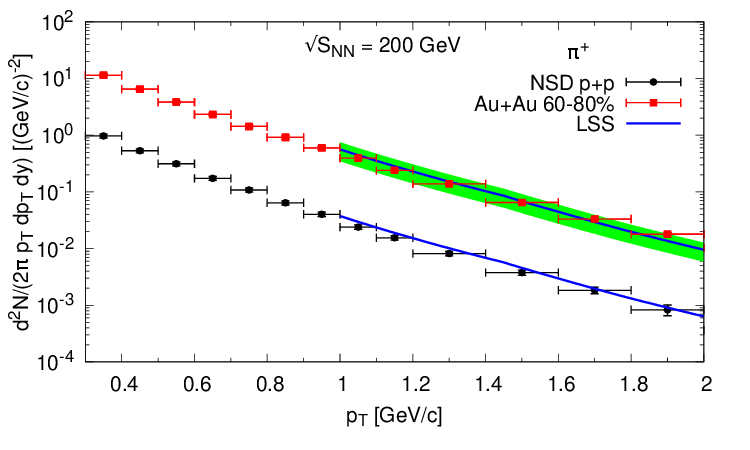}
\caption{
Transverse momentum spectra for $\pi^+$ in $p+p$ and Au$+$Au the most peripheral
collisions (centrality $60-80\%$) at $\sqrt{S} = 200\,{\rm GeV}$ and
midrapidity $|y|<0.5$. NLO pQCD reference for LSS parametrization of FFs
is scaled up for Au$+$Au by $N_{\rm coll}=21.2^{+6.6}_{\! -7.9}$,
Eqs.~(\ref{scal2}) and (\ref{scal3}), where $\s^{pp}_{inel}=42\,{\rm mb}$ for
$\sqrt{S} = 200\,{\rm GeV}$ \cite{TOTEM:2017asr}.
The error bands correspond to the systematic uncertainties of $N_{\rm coll}$.
Right panel: same as in the left panel but for NICA kinematics of
$p_T<2\,{\rm GeV}/c$.
}
\label{nfig1}
\end{figure}
\begin{figure}[h!]
\centering
\includegraphics[width=0.45\textwidth]{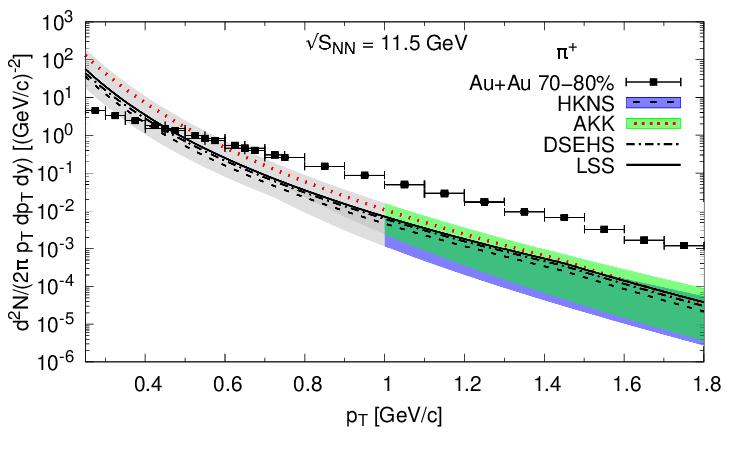}
\hfill
\includegraphics[width=0.45\textwidth]{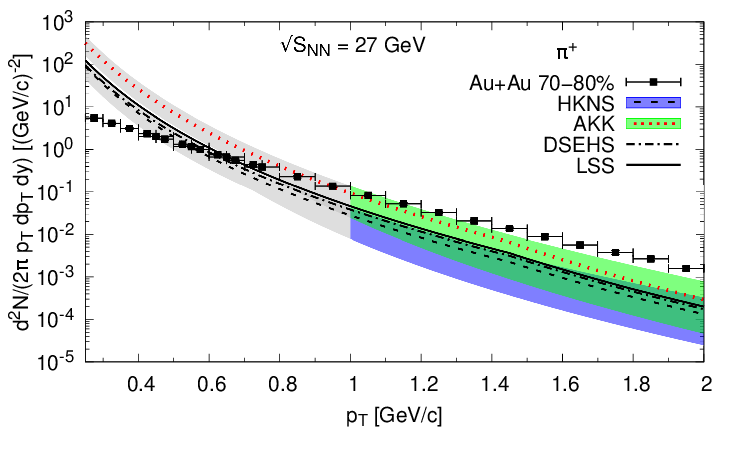}
\caption{
Transverse momentum spectra for $\pi^+$ in Au$+$Au the most peripheral
collisions (centrality $70-80\%$) at $\sqrt{S} = 11.5\,{\rm GeV}$ (left)
and $\sqrt{S} = 27\,{\rm GeV}$ (right), and midrapidity $|y|<0.1$,
\cite{STAR:2017sal}.
NLO pQCD references for different FFs parametrizations
are scaled up by $N_{\rm coll}=14\pm 7$, Eqs.~(\ref{scal2}) and (\ref{scal3}),
where $\s^{pp}_{inel}=33\,{\rm mb}$ for $\sqrt{S} = 27\,{\rm GeV}$ and
$31\,{\rm mb}$ for $\sqrt{S} = 11.5\,{\rm GeV}$ \cite{TOTEM:2017asr}.
The error bands, which for better visibility are shown only for AKK and HKNS
fits, correspond to the systematic uncertainties of $N_{\rm coll}$ and 
theoretical uncertainties due to $Q$ scale variation,
$p_T/2\leqslant Q \leqslant 2p_T$. Gray bands correspond to the non-perturbative
$p_T<1\,{\rm GeV}/c$ region where the pQCD analysis may be insufficient.
}
\label{nfig2}
\end{figure}
\begin{figure}[h!]
\centering
\includegraphics[width=0.45\textwidth]{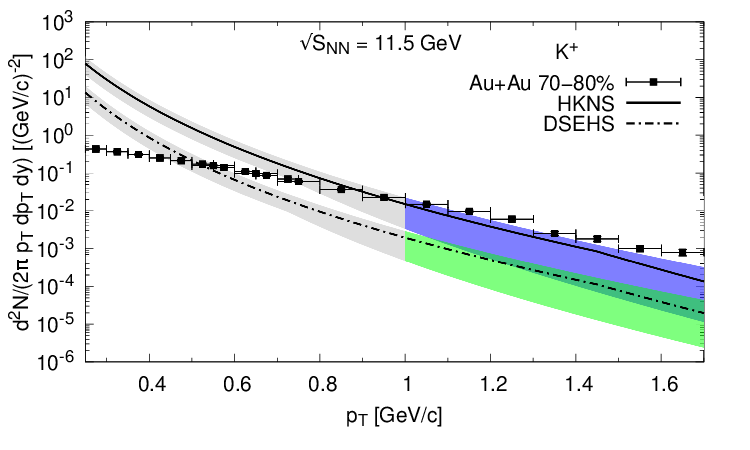}
\hfill
\includegraphics[width=0.45\textwidth]{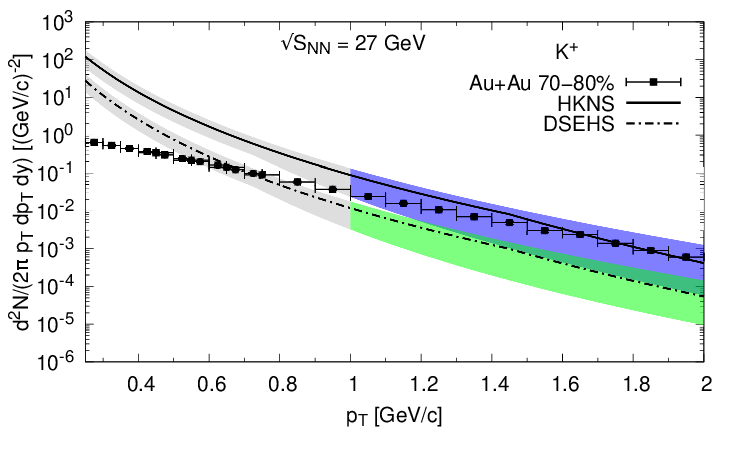}
\caption{
Same as in Fig.~\ref{nfig2} but for $K^+$.
}
\label{nfig3}
\end{figure}
\begin{figure}[h!]
\centering
\includegraphics[width=0.45\textwidth]{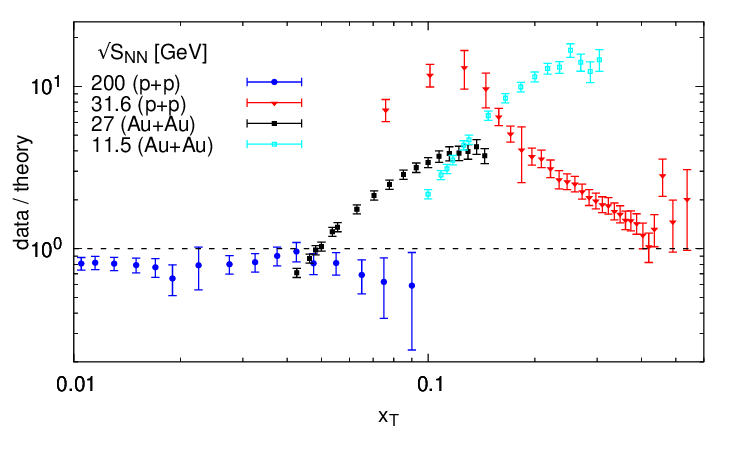}
\hfill
\includegraphics[width=0.45\textwidth]{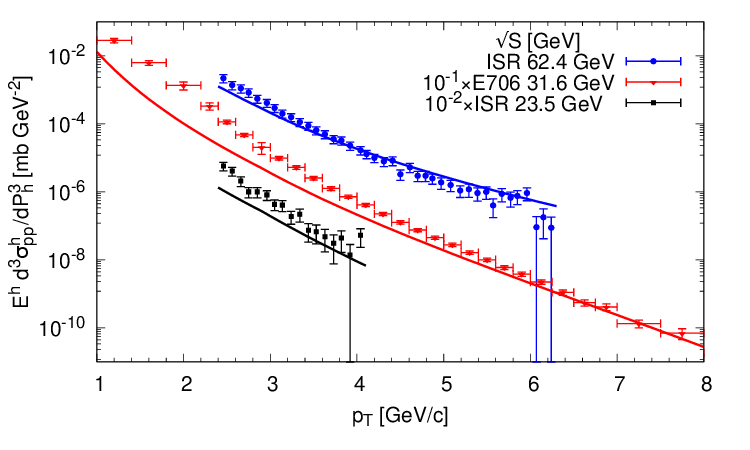}
\caption{
Left: ratios of data to theory for the inclusive $\pi^+$ (STAR), and also
$\pi^0$ (E706), production in $p+p$ and the most peripheral Au+Au collisions
as a function of $x_T$ at different $\sqrt{S_{NN}}$: 200, 27 and 11.5 GeV
at STAR, and 31.6 GeV at E706.  
Right: the NLO pQCD results for the $p_T$ spectra of $\pi^0$ at different
$\sqrt{S}$ compared to the measurements in $p+p$ collisions at ISR and E706.
}
\label{nfig4}
\end{figure}

One can expect a similarly good scaling behaviour of the $p_T$ spectra for
pions and kaons also at NICA kinematics.
Figs.~\ref{nfig2} and \ref{nfig3} present our results for $\pi^+$ and $K^+$
based on different FFs fits compared with the STAR data on Au$+$Au most
peripheral collisions at centrality $70-80\%$ and c.m. energies $\sqrt{S} = 11.5$
and 27 GeV \cite{STAR:2017sal}.
In our analysis for low energies and low $p_T<1\,{\rm GeV}/c$, we freeze QCD
evolution of PDFs and FFs at the scale $Q_0^2=1\,{\rm GeV^2}$. Nevertheless,
the pQCD predictions without non-perturbative transverse momentum corrections
(e.g. higher-twist effects in PDFs and FFs), which still are not under
theoretical control, makes this analysis insufficient.
Therefore, we draw attention mostly to our predictions for NICA obtained for
$1 < p_T < 2\, {\rm GeV}/c$.

In turn, for $p_T>1\,{\rm GeV}/c$, one sees from Figs.~\ref{nfig2} and
\ref{nfig3} that the NLO pQCD results for LSS and DSEHS FFs, which provide
the best fits to the STAR data at $\sqrt{S} = 200\,{\rm GeV}$, here, for much
lower energy, predict yields that are significantly below the data at low
$p_T$.
Such discrepancies have been observed or expected in all hard processes and
interpreted as an impact of an additional soft-gluon emission at the initial
state \cite{Apanasevich:2000eq,FermilabE706:2002wtp}.
This effect causes sizable parton $k_T$ which is not taken into account in
the NLO pQCD calculation.

Deviations between measured inclusive pion cross sections and NLO pQCD
calculations are visible mostly for $x_T = 2p_T/\sqrt{S} \gtrsim 0.1$
\cite{Apanasevich:2000eq}. This is illustrated in Fig.~\ref{nfig4} (left),
where we plot ratios of data to theory for the inclusive pion production in
$p+p$ and the most peripheral $Au+Au$ collisions as a function of $x_T$ at
different $\sqrt{S_{NN}}$.
In the right panel, we compare the $p_T$ spectra for pions at different
$\sqrt{S}$ to the measurements in $p+p$ collisions.
We use the LSS theoretical predictions and the experimental data on $\pi^+$,
and also $\pi^0$, at the energy scales from 11.5 up to 200 GeV
\cite{STAR:2017sal,Busser:1976su,FermilabE706:2002wtp,STAR:2006xud,
STAR:2006nmo,Agakishiev:2011dc,STAR:2006uve}.
It's seen that discrepancies between data and theory grow with decreasing
energy and become significant at about $\sqrt{S} = 60\,{\rm GeV}$.
Thus, scaling, Eqs.~(\ref{scal2}) and (\ref{scal3}), works well at
high energies, significantly above the energy scales specific for NICA.

Finally, let us discuss different behaviour of the antiparticle to particle
ratios of the $p_T$ spectra at the lower BES energies in comparison to those at
$\sqrt{S} = 200\,{\rm GeV}$. It could be a key to distinguish between
various FFs parametrizations.

The experimental results from Au+Au collisions at $\sqrt{S_{NN}} = 11.5$ and
27 GeV show that the $\pi^-/\pi^+$ ratio is mostly positive and close to
unity in contrast to our theoretical prediction, and also to the data at
$\sqrt{S} = 200\,{\rm GeV}$, where $\pi^+>\pi^-$, see Fig.~(\ref{nfig5}).
In turn, the ratio for kaons, $K^-/K+$, exhibit an interesting
trend, namely, it increases with increasing energy and lies significantly below
unity at the lower BES energies, see Fig.~(\ref{nfig6}). This reflects
the increasing contribution to $K^-$ production via pair production which
dominates over so called associated production at higher energies
(e.g. $NN\rightarrow KYN$, where $Y$ denotes a hyperon). These behaviours of
$\pi^-/\pi^+$ and $K^-/K+$ ratios are very little centrality dependent and
occur also for the peripheral collisions under study.

It is reasonable to assume that at a given energy the additional initial
state soft-gluon emission corrections are the same for particles and antiparticles
produced in $p+p$ collisions. Therefore, these corrections, which are not
included in the NLO pQCD analysis, can possibly be neglected when we study
the ratio of the $p_T$ spectra, $h^-/h^+$. In the light of this assumption,
one can see from Figs.~(\ref{nfig5}) and (\ref{nfig6}) that while the data
do not agree with theory there is some preference for the LSS-15 and DSEHS-14 FFs
for pions and DSEHS-17 for kaons over the HKNS one. 

For the difference cross sections, $\s_{pp}^{h^+ - h^-}$, the gap between
predictions for different FFs is undetectably small within the theoretical
uncertainty bands, see Fig.~(\ref{nfig7}), and the measurements of
$\sigma^{h^+ -h^-}$ cannot distinguish between various FFs sets.
\begin{figure}[h!]
\centering
\includegraphics[width=0.45\textwidth]{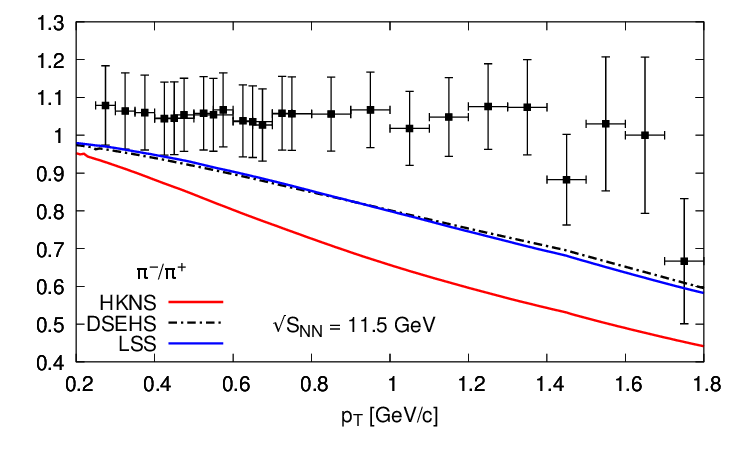}
\hfill
\includegraphics[width=0.45\textwidth]{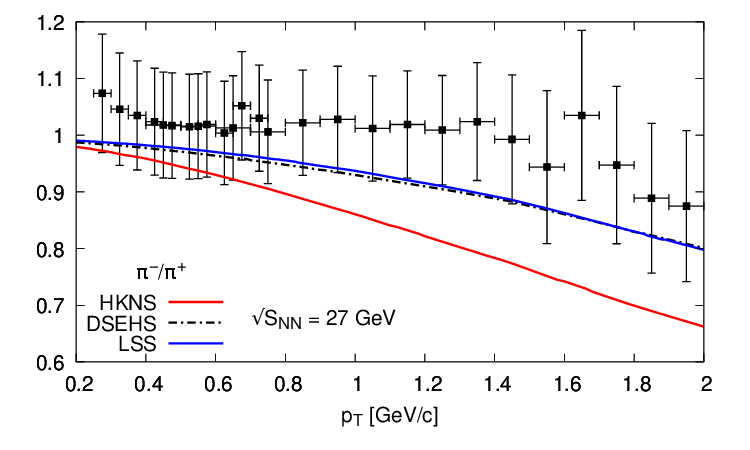}
\caption{
The ratios of the $p_T$ spectra for $\pi^-/\pi^+$ in Au$+$Au the most peripheral
collisions at $\sqrt{S} = 11.5\,{\rm GeV}$ (left)
and $\sqrt{S} = 27\,{\rm GeV}$ (right) compared with the NLO pQCD results.
}
\label{nfig5}
\end{figure}
\begin{figure}[h!]
\centering
\includegraphics[width=0.45\textwidth]{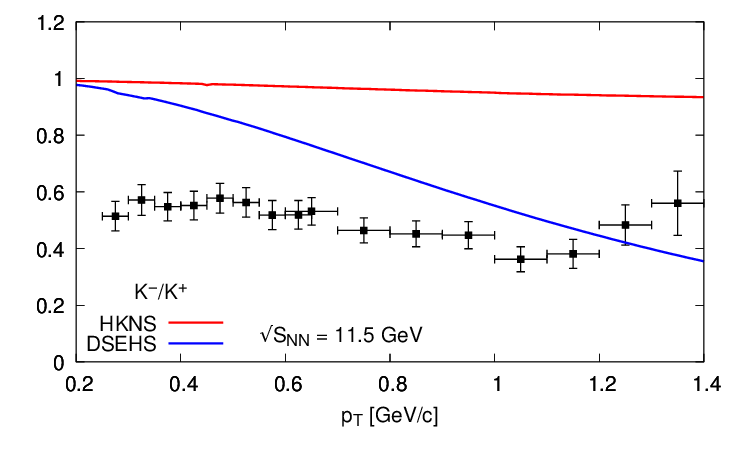}
\hfill
\includegraphics[width=0.45\textwidth]{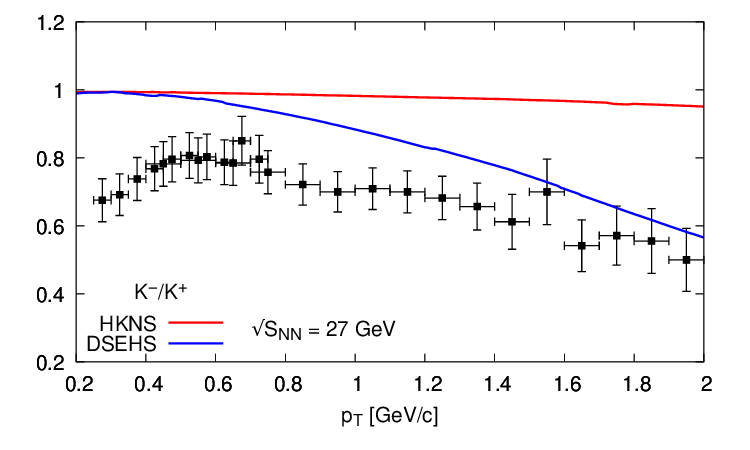}
\caption{
Same as in Fig.~\ref{nfig5} but for kaons.
}
\label{nfig6}
\end{figure}
\begin{figure}[h!]
\centering
\includegraphics[width=0.45\textwidth]{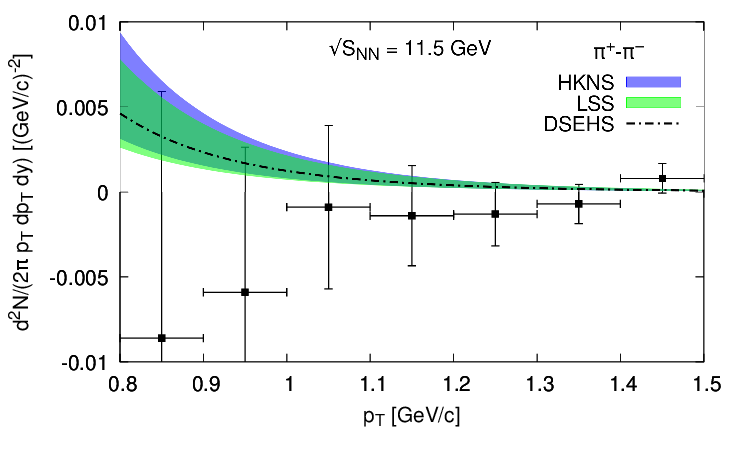}
\hfill
\includegraphics[width=0.45\textwidth]{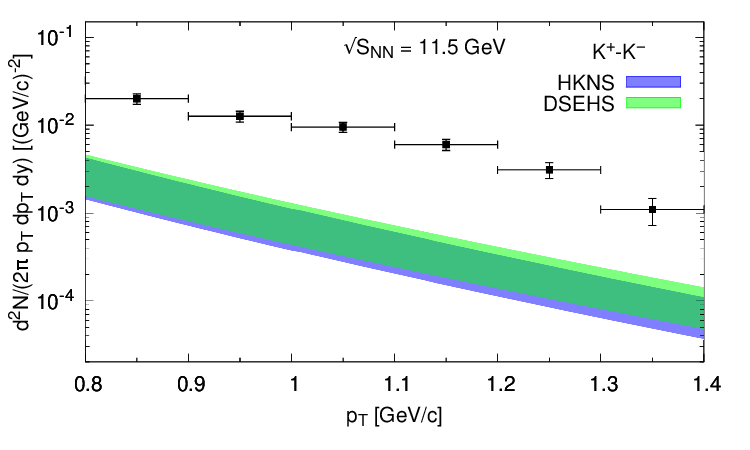}
\caption{
The $p_T$ spectra of the difference cross sections, $\pi^+-\pi^-$ (left),
and $K^+-K^-$ (right) in Au$+$Au the most peripheral collisions at
$\sqrt{S} = 11.5\,{\rm GeV}$ compared with the NLO pQCD results for
different FFs.
}
\label{nfig7}
\end{figure}

\section*{Conclusions}
There exist in the literature several different relatively up to date expressions for the pion and kaon
fragmentation functions: HKNS-07 \cite{Hirai:2007cx}, AKK-08 \cite{Albino:2008fy},
DSEHS-14~\cite{deFlorian:2014xna}, DSEHS-17~\cite{deFlorian:2017lwf} and LSS-15 \cite{Leader:2015hna}
and we have attempted, in this paper, to examine which, if any, of these is compatible with the STAR
data on semi-inclusive pion and kaon production in proton-proton collisions. We have thus compared
the predictions for the semi-inclusive cross-sections $pp \rightarrow \textrm{pions} + X$ and
$pp \rightarrow \textrm{kaons} + X$ based on these FFs with the data from the STAR collaboration.
Of the published FFs all except AKK respect isospin invariance.
As seen in Fig.~\ref{figu11}, AKK fails badly to describe the data on $pp \rightarrow K^+ + K^- - 2K_s^0$,
so is no longer considered a viable FF for kaons. All the others provide a good fit to the STAR data,
with both LSS-15 and DSEHS-14 for pions and DSEHS-17 for kaons significantly better than the others.

Furthermore, we have compared the results for the cross sections obtained within LO and NLO
approaches and studied their scale-$Q^2$ dependence, where $Q$ may vary from $p_T/2$ to $2p_T$.
We found that the NLO predictions fit clearly better the data, specially in the small $p_T$
range, and are less scale dependent than the LO ones.

In our paper, we have also obtained estimates of various semi-inclusive cross-sections
based on the above viable FFs for the future run of the NICA accelerator.
This is achieved comparing our NLO pQCD results with the existing data from the
STAR experiment in the BES program at RHIC on the hadron production in the most
peripheral
Au$+$Au collisions, where the nuclear effects are nefligible, at the energies similar
to those planned at NICA \cite{STAR:2017sal}.
Disagreement between data and theory at $\sqrt{S} = 11.5$ and 27~GeV shows
that a purely pQCD approach is inadequate and suggests the necessity to take into account
also higher-order effects of initial-state soft-gluon radiation,
\cite{Apanasevich:2000eq,FermilabE706:2002wtp}, not included in the NLO pQCD approach.
Nevertheless, these data on the $p_T$ spectra of $\pi^+$, $K^+$ and also the ratios
$\pi^-/\pi^+$ and $K^-/K^+$, (see Figs.~\ref{nfig2}, \ref{nfig3} and
\ref{nfig5}, \ref{nfig6}) seem favour LSS-15 and DSEHS-14 FFs for pions
and DSEHS-17 for kaons, similarly as at the energy scale $\sqrt{S} = 200\,{\rm GeV}$.
We have also check that the measurements of the difference cross sections,
$\sigma^{h^+ -h^-}$, cannot distinguish between various FFs sets (see Fig.~\ref{nfig7}).

\section*{Acknowledgements}
E.C. and D.K. acknowledge the support of the INRNE-BAS (Bulgaria) - JINR (Russia) collaborative Grant,
E.C. -- of Grant KP-06-N58/5 of the Bulgarian Science Foundation and D.K. -- of the Bogoliubov-Infeld Program.
D.K. is grateful to S.V. Mikhailov and A. Kotlorz for useful discussions on numerical analysis.


\appendix
\section{The partonic subprocesses contributing to $pp\to hX$}

Following the paper \cite{Owens:1977sj}, we obtain the expression for the inclusive
cross section $pp\to hX$:
\be
E^h\frac{d\sigma_{pp}^{h}}{d^3P^h} =\frac{1}{\pi}\,\sum_{ab\to cd}\int^1_{x_{a,min}}dx_a
\int^1_{x_{b,min}}dx_b\, \frac{1}{z }\,
q_a(x_a)q_b(x_b)\,\frac{d\hat\sigma_{ab}^{cd}}{dt}\,D_c^h(z )\label{A1}\,.
\ee
where
\be
 s=x_ax_b\,S,\quad  t=\frac{x_a}{z}\,T,\quad u=\frac{x_b}{z}\,U,\quad z =-\,\frac{x_a\,T+x_b\,U}{x_a\,x_b\,S}\label{Mand}
\ee
The lower limits of integration are determined by the conditions $z <1$ and $s+t+u=0$
\cite{Owens:1977sj}:
\be
x_{a,min}&=&\frac{-U}{T+S} \\
x_{b,min}&=&\frac{-x_aT}{x_aS+U}
\ee
Hence, we can calculate the contributions from the different partonic
subprocesses of Eq.~(\ref{A1}),
\be
q_a(x_a)q_b(x_b)\,\frac{d\hat\sigma_{ab}^{cd}}{dt}\,D_c^h(z )\label{A2}\, .
\ee
Below, we give formulas for all possible $ab\to cd$ contributions together
with detailed examples.

The general formula  is:
\be
q_a(x_a)q_b(x_b)\,\left[\frac{d\hat\sigma_{ab}^{cd}}{dt}\,D_c^h(z )+\frac{d\hat\sigma_{ab}^{cd}}{du}\,D_d^h(z )\right]+q_a(x_b)q_b(x_a)\,\left[\frac{d\hat\sigma_{ab}^{cd}}{du}\,D_c^h(z )+\frac{d\hat\sigma_{ab}^{cd}}{dt}\,D_d^h(z )\right],
\label{K2}
\ee
where
\begin{enumerate}
\item $\hat\sigma_1$: $q_iq_j\to q_iq_j,\quad \bar q_i\bar q_j\to \bar q_i \bar q_j,\quad q_i\bar q_j\to q_i\bar q_j,
\quad \bar q_i q_j \to \bar q_i q_j,\quad i\neq j$
\be
\hat\sigma_1:~~~~~~
\sum_{i\neq j}&& q_i(x_a)q_j(x_b)\left[D_{q_i}^h\,\frac{d\hat\sigma_1}{dt}+D_{q_j}^h\,\frac{d\hat\sigma_1}{du}\right] +
\bar q_i(x_a)\bar q_j(x_b)\left[D_{\bar q_i}^h\,\frac{d\hat\sigma_1}{dt}+D_{\bar q_j}^h\,\frac{d\hat\sigma_1}{du}\right]\nn
&& +\, q_i(x_a)\bar q_j(x_b)\left[D_{q_i}^h\,\frac{d\hat\sigma_1}{dt}+D_{\bar q_j}^h\,\frac{d\hat\sigma_1}{du}\right] +\bar q_i(x_a)q_j(x_b)\left[D_{\bar q_i}^h\,\frac{d\hat\sigma_1}{dt}+D_{q_j}^h\,\frac{d\hat\sigma_1}{du}\right] \nn
=\sum_{i\neq j}&&\left[ q_i(x_a)D_{q_i}^h+\bar q_i(x_a)D_{\bar q_i}^h\right]\,
 \wt q_j(x_b)\,\frac{d\hat\sigma_1}{dt} +
\left[ q_j(x_b)D_{q_j}^h+\bar q_j(x_b)D_{\bar q_j}^h\right]\,\tilde  q_i(x_a)\,
\frac{d\hat\sigma_1}{du}\nn
 &&i\neq j: \,i=u, j=d,s;\,\, i=d,j=u,s,\quad i=s,j=u,d
\label{sig1}\, .
\ee
\be
\sigma_1^{h-\bar h}:~~~~~
\sum_{i\neq j}&&\left\{q_{iV}(x_a)\,\wt q_j(x_b)\,D_{q_{iV}}^h\,\frac{d\hat\sigma_1}{dt}+
 q_{jV}(x_b)\,\wt q_i(x_a)\,D_{q_{jV}}^h\,
\frac{d\hat\sigma_1}{du}\right\},\nn
 &&i\neq j: \,i=u, j=d,s;\,\, i=d,j=u,s,\quad i=s,j=u,d
\label{sig1b}\, .
\ee
\be
\sigma_1^{h+\bar h}:~~~~~
=\sum_{i\neq j}&&\left[ \wt q_i(x_a)\wt q_j(x_b)\, D_{ q_i}^{h+\bar h}\,
  \,\frac{d\hat\sigma_1}{dt} +
\wt q_j(x_b)\,\wt  q_i(x_a)\,D_{ q_j}^{h+\bar h}\,
\frac{d\hat\sigma_1}{du}\right]\nn
 &&i\neq j: \,i=u, j=d,s;\,\, i=d,j=u,s,\quad i=s,j=u,d
\label{sig1aa}\, .
\ee

\item $\hat\sigma_2$: $q_iq_i\to q_iq_i,\quad \bar q_i\bar q_i\to \bar q_i \bar q_i$
\be
\hat\sigma_2:~~~~~~
\sum_{i=u,d,s}\left[ q_i(x_a)q_i(x_b)\,\,D_{q_i}^h(z )+\bar q_i(x_a)\bar q_i(x_b)\, D_{\bar q_i}^h(z )\right]
\frac{d\hat\sigma_2}{dt}
\label{sig2}\, ,
\ee
symmetric in $t-u$.
\be
\hat\sigma_2^{h-\bar h}:~~~~~~
&&\sum_{q=u,d,s}\left[ q(x_a)q(x_b)\,-\bar q(x_a)\bar q(x_b)\right]\, D_{q_V}^h(z )
\frac{d\hat\sigma_2}{dt}\nn
&=&\sum_{q=u,d,s}\frac{1}{2}\left[ q_V(x_a)\wt q(x_b)\,+ \wt q(x_a)\, q_V(x_b)\right]\, D_{q_V}^h(z )
\frac{d\hat\sigma_2}{dt}
\label{sig2b}\, .
\ee
\be
\hat\sigma_2^{h+\bar h}:~~~~~~
&&\sum_{q=u,d,s}\,\left[\, q(x_a)q(x_b)\,+\bar q(x_a)\bar q(x_b)\,\right]\,
\frac{d\hat\sigma_2}{dt}\,D_{q}^{h+\bar h}(z )
\label{sig2bb}\, .
\ee

\item $\hat\sigma_3$: $q_i\bar q_i\to q_j\bar q_j,\quad i\neq j$
\be
\hat\sigma_3:~~~~~~
\sum_{i\neq j}
\left [\, q_i(x_a)\bar{q}_i(x_b)
+ \bar{q}_i(x_a)q_i(x_b)\,\right]\,\frac{d\hat\sigma_3}{dt}\,D_{j+\bar j}^h(z )
\label{sig3}\, .
\ee
symmetric in $t-u$.
\be
\hat\sigma_3^{h+\bar h}:~~~~~~
\sum_{i\neq j}
2\,\left [\, q_i(x_a)\bar{q}_i(x_b)
+ \bar{q}_i(x_a)q_i(x_b)\,\right]\,\frac{d\hat\sigma_3}{dt}\,D_{j}^{h+\bar h}(z )
\label{sig3aa}\, .
\ee
\item $\hat\sigma_4$: $q_i\bar q_i\to q_i\bar q_i$:
\be
\hat\sigma_4:~~~~~~
\sum_{i} \left\{
q_i(x_a)\bar{q}_i(x_b)\,\left(\frac{d\hat\sigma_4}{dt}\,D_{q_i}^h(z )+
\frac{d\hat\sigma_4}{du}\,D_{\bar{q_i}}^h(z )\right)
 +\,\bar{q}_i(x_a)q_i(x_b)\,\left(\frac{d\hat\sigma_4}{dt}\,D_{\bar{q}_i}^h(z )+
\frac{d\hat\sigma_4}{du}\,D_{q_i}^h(z )\right) \right \}\nn
\label{sig4}\, .
\ee
\be
\hat\sigma_4^{h-\bar h}:~~~~~~
&&\sum_{q=u,d,s}
\left[q(x_a)\bar{q}(x_b)\,-
 \,\bar{q}(x_a)q(x_b)\right]\,\left(\frac{d\hat\sigma_4}{dt}-
\frac{d\hat\sigma_4}{du}\right)\,D_{q_V}^h(z ) \nn
&&=\sum_{q=u,d,s}
\left[q_V(x_a){q}(x_b)\,-
 \,{q}(x_a)q_V(x_b)\right]\,\left(\frac{d\hat\sigma_4}{dt}-
\frac{d\hat\sigma_4}{du}\right)\,D_{q_V}^h(z )\nn
&&=\sum_{q=u,d,s}
\left[q_V(x_a)\bar{q}(x_b)\,-
 \,\bar{q}(x_a)q_V(x_b)\right]\,\left(\frac{d\hat\sigma_4}{dt}-
\frac{d\hat\sigma_4}{du}\right)\,D_{q_V}^h(z ) \nn
&&=\sum_{q=u,d,s}
\frac{1}{2}\left[q_V(x_a)\tilde{q}(x_b)\,-
 \,\tilde{q}(x_a)q_V(x_b)\right]\,\left(\frac{d\hat\sigma_4}{dt}-
\frac{d\hat\sigma_4}{du}\right)\,D_{q_V}^h(z )
\label{sig4a}\, .
\ee
\be
\hat\sigma_4^{h+\bar h}:~~~~~~
&&\sum_{q=u,d,s}
\left[q(x_a)\bar{q}(x_b)\,+
 \,\bar{q}(x_a)q(x_b)\right]\,\left(\frac{d\hat\sigma_4}{dt}+
\frac{d\hat\sigma_4}{du}\right)\,D_{q}^{h+\bar h}(z ) \label{sig4aa}\
\ee

\item $\hat\sigma_5$: $q_i\bar q_i\to gg$
\be
\hat\sigma_5:~~~~~~
\sum_{i=u,d,s}\left[\,
q_i(x_a)\bar{q}_i(x_b)+q_i(x_b)\bar q_i(x_a)\,\right]\frac{d\hat\sigma_5}{dt}\,D_{g}^h(z )
\label{sig5}\, ,
\ee
symmetric in $t-u$.
\be
\hat\sigma_5^{h-\bar h} =0
\ee
\be
\hat\sigma_5^{h+\bar h}:~~~~~~
\sum_{i=u,d,s}2\left[\,
q_i(x_a)\bar{q}_i(x_b)+q_i(x_b)\bar q_i(x_a)\,\right]\,\frac{d\hat\sigma_5}{dt}\,D_{g}^h(z )
\label{sig5aa}\, ,
\ee

\item $\hat\sigma_6$: $ gg\to q_i\bar q_i$
\be
\hat\sigma_6:~~~~~~
g(x_a)g(x_b)\,\frac{d\hat\sigma_6}{dt}\sum_{i=u,d,s}
\left [ D_{q_i}^h(z ) + D_{\bar{q}_i}^h(z )\right ]
\label{sig6}\, ,
\ee
symmetric in $t-u$.
\be
\hat\sigma_6^{h-\bar h} =0
\ee
\be
\hat\sigma_6^{h+\bar h}:~~~~~~
2\,g(x_a)g(x_b)\,\frac{d\hat\sigma_6}{dt}
\, \sum_{i=u,d,s}D_{q_i}^{h+\bar h}(z )
\label{sig6aa}\, ,
\ee
where we have used that C-inv. implies $D_g^h=D_g^{\bar h}$

\item $\hat\sigma_7$: $q_ig\to q_i g$

\be
\hat\sigma_7:~~&&
\sum_{i}\left\{\left[
\left(q_i(x_a)D_{q_i}^h+\bar q_i(x_a)D_{\bar q_i}^h\right)\,g(x_b)\frac{d\hat\s_7}{dt}+\left(q_i(x_a)+
 \bar q_i(x_a)\right)g(x_b)\,D_g^h\,\frac{d\hat\sigma_7}{du}\right]\right.\\
 &&\hspace*{0.5cm}+\left.\left[\left(q_i(x_b)D_{q_i}^h+\bar q_i(x_b)D_{\bar q_i}^h\right)\,g(x_a)\frac{d\hat\s_7}{du}+\left(q_i(x_b)+\bar q_i(x_b)\right)g(x_a)\,D_g^h\,\frac{d\hat\s_7}{dt}\right]
\right \}
\label{sig7}\, .
\ee
\be
\hat\sigma_7^{h-\bar h}:~~&&
\sum_{q=u,d,s}\left\{
q_V(x_a)\,g(x_b)\frac{d\hat\s_7}{dt}+q_V(x_b)\,g(x_a)\frac{d\hat\s_7}{du}
\right \}D_{q_V}^h
\label{sig7a}\, .
\ee
\be
\hat\sigma_7^{h+\bar h}:~~&&
\sum_{q=u,d,s}\left\{\left[
\wt q(x_a)\,g(x_b)\frac{d\hat\s_7}{dt}+\wt q(x_b)\,g(x_a)\frac{d\hat\s_7}{du}
\right ] D_{q}^{h+\bar h}\right. \nn
&&\hspace*{1.2cm}+ \left. 2\left[ \wt q(x_a)\,g(x_b)\frac{d\hat\s_7}{du}+\wt q(x_b)\,g(x_a)\frac{d\hat\s_7}{dt}
\right ] D_{g}^{h}\right\}
\label{sig7aa}\, .
\ee

\item $\hat\sigma_8$: $gg\to gg$
\be
\hat\sigma_8:~~~~~~
g(x_a)g(x_b)\,\frac{d\hat\sigma_8}{dt}\,D_g^h(z )
\label{sig8}\, ,
\ee
symmetric in $t-u$.
\be
\hat\sigma_8^{h-\bar h} =0
\ee
\be
\hat\sigma_8^{h+\bar h}:~~~~~~
2\,g(x_a)g(x_b)\,\frac{d\hat\sigma_8}{dt}\,D_g^h(z )
\label{sig8aa}\, ,
\ee
\end{enumerate}

The  partonic cross sections that enter the difference cross sections, averaged over initial and summed over final spin and colour,
are \cite{Owens:1977sj}:
\be
\frac{d\hat\sigma_i(ab\to cd)}{dt}&=&\frac{\pi \a_s^2(Q^2)}{s^2}\,\vert M_i (s,t,u)\vert^2
\ee
where the matrix elements $\vert M_i (s,t,u)\vert^2$ are:
\be
\vert M_1 (s,t,u)\vert^2&=&\,\frac{4}{9}\,\frac{ s^2 + u^2}{ t^2}\\
 \vert M_2 (s,t,u)\vert^2&=&
 \,\frac{4}{9}\,\left(\frac{ s^2 + u^2}{ t^2}+\frac{ s^2 + t^2}{ u^2}\right)
-\frac{8}{27}\,\frac{ s^2}{ t u}\\
\vert M_4 (s,t,u)\vert^2&=&
\,\frac{4}{9}\,\left(\frac{ s^2 + u^2}{ t^2}+\frac{ u^2 + t^2}{ s^2}\right)
-\frac{8}{27}\,\frac{ u^2}{ st }\\
\vert M_7 (s,t,u)\vert^2&=&\,-\frac{4}{9}\,\frac{ s^2 + u^2}{ u s}+
\,\frac{ s^2 + u^2}{ t^2}\cdot
\ee

\bibliography{xref}
\bibliographystyle{apsrev}

\end{document}